\documentclass[12pt]{article}

\usepackage{times,epsfig,graphics,amssymb,latexsym,amsmath,setspace}
\usepackage{array,threeparttable,hyperref,url}
\usepackage[usenames]{color}
\usepackage[authoryear,sort]{natbib}
\usepackage{paralist}
\usepackage{mathabx}
\usepackage{algorithm,algorithmic}
\usepackage{ntheorem}
\usepackage[normalem]{ulem}
\usepackage{authblk}
\usepackage[symbol*]{footmisc}
\usepackage{tabularx,ragged2e}
\newtheorem{remark}{Remark}

\allowdisplaybreaks

\newcolumntype{L}[1]{>{\raggedright\arraybackslash}p{#1}}
\newcolumntype{C}[1]{>{\centering\arraybackslash}p{#1}}
\newcolumntype{R}[1]{>{\raggedleft\arraybackslash}p{#1}}

\DefineFNsymbolsTM{myfnsymbols}{
  \textasteriskcentered *
  \textdagger    \dagger
  \textdaggerdbl \ddagger
  \textsection   \mathsection
  \textbardbl    \|%
  \textparagraph \mathparagraph
}%
\setfnsymbol{myfnsymbols}

\theoremstyle{empty}

\theoremstyle{plain}

\def\wt{\widetilde}

\def\wh{\widehat}

\def\log{\hbox{log}}

\def\boxit#1{\vbox{\hrule\hbox{\vrule\kern6pt
          \vbox{\kern6pt#1\kern6pt}\kern6pt\vrule}\hrule}}

\def\bse{\begin{eqnarray*}}
\def\ese{\end{eqnarray*}}
\def\be{\begin{eqnarray}}
\def\ee{\end{eqnarray}}
\def\bq{\begin{equation}}
\def\eq{\end{equation}}

\def\wh{\widehat}

\def\trans{^{\rm T}}

\newtheorem{proposition}{Proposition}

\newcommand{\blem}{\begin{lemma}}
\newcommand{\elem}{\end{lemma}}
\newcommand{\bthe}{\begin{theorem}}
\newcommand{\ethe}{\end{theorem}}

\newtheorem{lemma}{Lemma}
\newtheorem{theorem}{Theorem}
\def\bfa{{\bf a}}
\def\bfA{{\bf A}}
\def\bfb{{\bf b}}
\def\bfB{{\bf B}}

\def\bfC{{\bf C}}
\def\bfd{{\bf d}}
\def\bfD{{\bf D}}
\def\bfe{{\bf e}}
\def\bfE{{\bf E}}

\def\bfg{{\bf g}}
\def\bfG{{\bf G}}

\def\bfI{{\bf I}}

\def\bfM{{\bf M}}

\def\bfN{{\bf N}}

\def\bfP{{\bf P}}

\def\bfR{{\bf R}}

\def\bfS{{\bf S}}

\def\bfT{{\bf T}}
\def\bfu{{\bf u}}

\def\bfv{{\bf v}}
\def\bfV{{\bf V}}

\def\bfx{{\bf x}}
\def\bfX{{\bf X}}
\def\bfy{{\bf y}}

\def\bfz{{\bf z}}

\def\blambda{{\mbox{\boldmath $\lambda$}}}
\def\bLambda{{\mbox{\boldmath $\Lambda$}}}
\def\bmu{{\mbox{\boldmath $\mu$}}}

\def\bpsi{{\mbox{\boldmath $\psi$}}}

\def\bphi{{\mbox{\boldmath $\phi$}}}
\def\bPhi{{\mbox{\boldmath $\Phi$}}}
\def\bepsilon{{\mbox{\boldmath $\epsilon$}}}
\def\bvarepsilon{{\mbox{\boldmath $\varepsilon$}}}

\def\bxi{{\mbox{\boldmath $\xi$}}}

\def\delete#1{\iffalse #1 \fi}

\def\bse{\begin{eqnarray*}}
\def\ese{\end{eqnarray*}}
\def\bee{\begin{enumerate}}
\def\eee{\end{enumerate}}
\def\bqe{\begin{eqnarray}}
\def\eqe{\end{eqnarray}}
\def\bed{\begin{description}}
\def\eed{\end{description}}
\def\bei{\begin{itemize}}
\def\eei{\end{itemize}}

\def\argmin{\mathop{\rm argmin}}

\def\real{{\mathbb R}}

\def\pmb#1{\setbox0=\hbox{#1}%
    \kern-.025em\copy0\kern-\wd0
    \kern.05em\copy0\kern-\wd0
    \kern-.025em\raise.0433em\box0 }
\def\pmbh#1#2{\setbox0=\hbox{#1}%
    \setbox1=\hbox{#2}%
    \kern-.025em\copy0\kern-\wd0
    \kern.05em\copy1\kern-\wd0
    \kern-.025em\raise.0433em\box0 }

\def\binom#1#2{{#1\choose #2}}
\def\frac#1#2{{#1\over#2}}

\def\boxit#1{\vbox{\hrule\hbox{\vrule\kern6pt
   \vbox{\kern6pt#1\kern6pt}\kern6pt\vrule}\hrule}}

\def\listing#1{\vskip 4mm\begin{verbatim}\input#1 \vskip 4mm}
\def\thick#1{\hbox{\rlap{$#1$}\kern0.25pt\rlap{$#1$}\kern0.25pt$#1$}}

\def\wt{\widetilde}
\def\wh{\widehat}

\def\bfa{{\bf a}}  \def\bfA{{\bf A}}
\def\bfb{{\bf b}}  \def\bfB{{\bf B}}
  \def\bfC{{\bf C}}
\def\bfd{{\bf d}}  \def\bfD{{\bf D}}
\def\bfe{{\bf e}}  \def\bfE{{\bf E}}
  
\def\bfg{{\bf g}}  \def\bfG{{\bf G}}
  
  \def\bfI{{\bf I}}

  \def\bfM{{\bf M}}
  \def\bfN{{\bf N}}
  
  \def\bfP{{\bf P}}
  
  \def\bfR{{\bf R}}
  \def\bfS{{\bf S}}
  \def\bfT{{\bf T}}
\def\bfu{{\bf u}}  
\def\bfv{{\bf v}}  \def\bfV{{\bf V}}
  
\def\bfx{{\bf x}}  \def\bfX{{\bf X}}
\def\bfy{{\bf y}}  
\def\bfz{{\bf z}}  
\def\bfzero{{\bf 0}}
\def\bfone{{\bf 1}}

\def\pmbh{{\pmb h}}

\def\calA{{\cal A}}
\def\calB{{\cal B}}

\def\calE{{\cal E}}

\def\calL{{\cal L}}

\def\calN{{\cal N}}

\def\calP{{\cal P}}

\renewcommand\today{\ifcase\month\or
   Jan\or Feb\or Mar\or Apr\or May\or
   Jun\or Jul\or Aug\or Sep\or Oct\or Nov\or
   Dec\fi
   \space\number\day, \number\year}

\def\boxit#1{\vbox{\hrule\hbox{\vrule\kern6pt\vbox{\kern6pt#1\kern6pt}\kern6pt\vrule}\hrule}}

\catcode`@=11
\def\@evenhead{\vbox{\hbox to \textwidth{\tiny \hfill \hfill \today } }}
\def\@oddhead{\vbox{\hbox to \textwidth{\tiny \hfill \hfill \today } }}

\def\argmin{\mathop{\rm argmin}}

\def\author.arg{
\medskip
Binhuan Wang$^{1}$\footnotemark, Yilong Zhang$^2$, Will Wei Sun$^3$, Yixin Fang$^4$\\
\medskip
{\it $^1$New York University School of Medicine}\\
{\it $^2$Merck Research Laboratories} \\
{\it $^3$University of Miami School of Business Administration} \\
{\it $^4$Department of Mathematical Sciences, New Jersey Institute of Technology}
}

\def\tit.arg{Sparse Convex Clustering}

\def\tits.arg{Supplementary Materials for: \\
Sparse Convex Clustering}

\begin{document}
\pagenumbering{arabic}
\setcounter{page}{1}
\baselineskip=14pt

\begin{center}
{\Large \tit.arg} \\

\vskip 3mm

\author.arg
\end{center}

\footnotetext
{ Correspondence to: 650 First Avenue Rm 578, New York, NY 10016; Email: Binhuan.Wang@nyumc.org}

\vskip 3mm

\date{}

\begin{abstract}
Convex clustering, a convex relaxation of k-means clustering and hierarchical
clustering, has drawn recent attentions since it nicely addresses
the instability issue of traditional non-convex clustering methods.
Although its computational and statistical properties have been recently
studied, the performance of convex clustering has not yet been investigated
in the high-dimensional clustering scenario, where the data contains a
large number of features and many of them carry no information about
the clustering structure. In this paper, we demonstrate that the performance
of convex clustering could be distorted when the uninformative features
are included in the clustering. To overcome it, we introduce a new
clustering method, referred to as \textit{Sparse Convex Clustering},
to simultaneously cluster observations and conduct feature selection.
The key idea is to formulate convex clustering in a form of regularization,
with an adaptive group-lasso penalty term on cluster centers. In order
to optimally balance the trade-off between the cluster fitting and
sparsity, a tuning criterion based on clustering stability is developed.
Theoretically, we obtain a finite sample error bound for our estimator and further establish its variable selection consistency.
The effectiveness of the proposed method is examined through a variety of numerical
experiments and a real data application.
\end{abstract}

\noindent

\doublespacing

\section{Introduction}

Cluster analysis is an unsupervised learning method and aims to assign
observations into a number of clusters such that observations in the
same group are similar to each other. Traditional clustering methods
such as k-means clustering, hierarchical clustering, and Gaussian mixture
models take a greedy approach and suffer from instabilities due to
their non-convex optimization formulations.

To overcome the instability issues of these traditional clustering methods, a new clustering algorithm, \textit{Convex Clustering},
has been recently proposed \citep{Pelckmans.ea:2005,Lindsten.ea:2004,Hocking2011}. Let
$\mathbf{X}\in\mathbb{R}^{n\times p}$ be a data matrix with $n$
observations $X_{i\cdot}$, $i=1,\cdots,n$, and $p$ features. Convex
clustering for these $n$ observations solves the following minimization
problem:
\begin{equation}
\min_{\mathbf{A}\in \mathbb{R}^{n\times p}}\frac{1}{2}\sum_{i=1}^{n}||X_{i\cdot}-A_{i\cdot}||_{2}^{2}+\gamma \sum_{i_{1}<i_{2}}||A_{i_{1}\cdot}-A_{i_{2}\cdot}||_{q},\label{eq:obj-cc}
\end{equation}
where $A_{i\cdot}$ is the $i$-th row of $\mathbf{A}$ and $\|\cdot\|_{q}$ is the $L_{q}$-norm of a vector with $q \in \{1,2,\infty\}$. Note that both k-means clustering and hierarchical
clustering consider $L_{0}$-norm in the second term, which leads to
a non-convex optimization problem \citep{Hocking2011,Tan2015}. Therefore, convex
clustering can be viewed as a convex relaxation of k-means clustering and hierarchical
clustering, and the convex relaxation ensures that it achieves a unique
global minimizer.

Due to the fused-lasso penalty \citep{Tibshirani.ea:2005} in the second term of (\ref{eq:obj-cc}),
the above formulation encourages that some of the rows of the solution $\widehat{\bfA}$
are identical. If $\widehat{A}_{i_{1}\cdot}=\widehat{A}_{i_{2}\cdot}$,
then observation $i_{1}$ and observation $i_{2}$ are said to belong
to the same cluster. The tuning parameter $\gamma$ in (\ref{eq:obj-cc})
controls the number of unique rows of $\widehat{\bfA}$, that is,
the number of estimated clusters. When $\gamma=0$, $\widehat{\bfA}={\bfX}$,
and therefore each observation by itself is a cluster. As $\gamma$
increases, some of the rows of $\widehat{\bfA}$ become identical,
which demonstrates a fusion process. For sufficiently large $\gamma$,
all the rows of $\widehat{\bfA}$ will be identical, implying that
all the observations are estimated to belong to a single cluster.
Compared to traditional non-convex clustering methods, the
solution $\widehat{\bfA}$ from convex clustering is unique for
each given $\gamma$ since the objective function in (\ref{eq:obj-cc})
is strictly convex.

In recent years, the computational and statistical properties of convex
clustering have been investigated. In particular, \citet{Zhu2014} provided conditions for convex clustering
to recover the true clusters, \citet{Chi2015} proposed efficient
and scalable implementations for convex clustering, and \citet{Tan2015}
studied several statistical properties of convex clustering. While convex clustering enjoys nice theoretical properties and is
computationally efficient, its performance can be severely deteriorated
when clustering high-dimensional data where the number of features
becomes large and many of them may contain no information about the
clustering structure. Our extensive experimental studies demonstrate
that in high-dimensional scenarios the performance of convex clustering
is unsatisfactory when the uninformative features are included in
the clustering. To overcome such a difficulty, a more appropriate convex
clustering algorithm that can simultaneously perform cluster analysis
and select informative variables is in demand.

In this article, we introduce a new clustering method, \textit{Sparse
Convex Clustering}, to incorporate the sparsity into convex clustering of high dimensional data. The key idea is to formulate convex clustering
in a form of regularization, with an adaptive group-lasso penalty
term on cluster centers to encourage the sparsity. Despite its simplicity,
this regularization operator demands more challenging computational
and statistical analysis than those in original convex clustering.
In particular, computationally, we need to reformulate the sparse convex
clustering into a few sub-optimization problems and then solve each
individual one via a pseudo regression formulation. To prove an unbiased
estimator for the degrees of freedom of the proposed sparse convex clustering
method, we need to carefully quantify the impact of variable selection
due to the group lasso penalty. Moreover, we provide a non-asymptotic analysis for the prediction error of our sparse convex clustering estimator. Under a high-dimensional scenario where the dimension diverges with the sample size, our estimator is further shown to be consistent in variable selection.
Note that our method is not only theoretical sound, but also practically promising. The superior performance
of our procedure is demonstrated in extensive simulated examples and
a real application of hand movement clustering.

We demonstrate the superior performance of the proposed method using a dataset generated from the fourth simulation setting in Section 5. In this dataset, there are 60 subjects from 4 clusters and 500 features, among which the first 20 features are informative. Figure \ref{heatmap} compares the performance of convex clustering (indicated as AMA on the left panel) and sparse convex clustering (indicated as S-AMA on the right panel), by visualizing the regularized feature matrix $\widehat{\bfA}$ estimated in each of the two methods. The heap maps show that sparse convex clustering screens out those uninformative features and therefore improves the clustering performance.

\begin{figure}[!htb]
\protect\caption{The heat maps of $p\times n$ matrix $\widehat{\bfA}\trans$, estimated from Convex clustering and Sparse Convex clustering, respectively. Data are generated using Simulation Setting 4 where there are 60 subjects in 4 clusters and 500 features, among which the first 20 features are informative.}
\centering \includegraphics[scale=0.4]{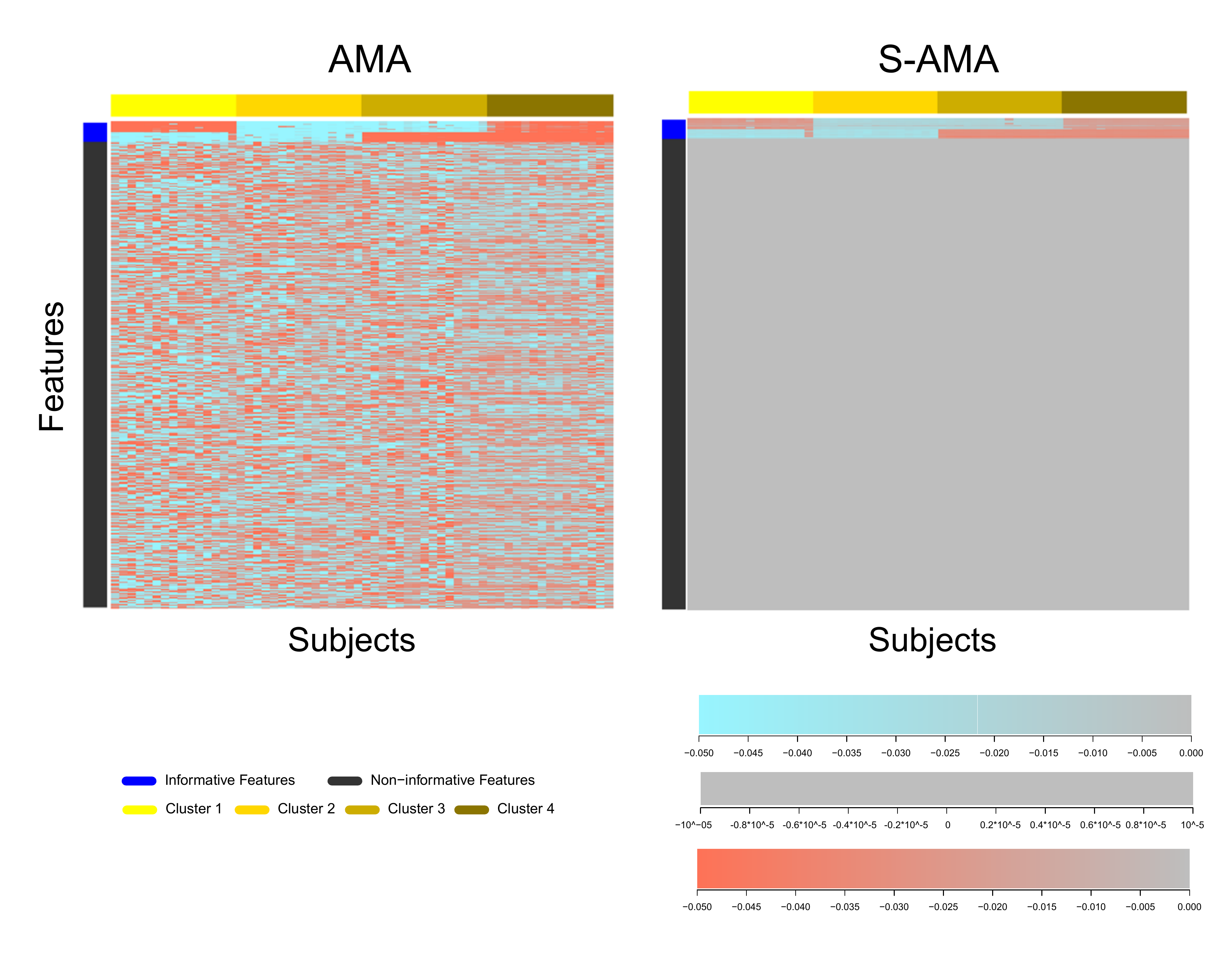}\label{heatmap}
\end{figure}

\subsection{Related Work}

A related paper on convex clustering is its efficient implementations proposed by \citet{Chi2015} and its extension to convex biclustering has been developed by \cite{Chi2016convex}.  Two efficient algorithms ADMM and AMA are introduced while they are mainly designed for clustering low-dimensional data. In order
to address high dimensionality, one key ingredient of our sparse convex
clustering method is a new regularization penalty built upon their
ADMM and AMA algorithms to encourage the sparsity structure of the
clustering centers. As will be shown in experimental studies, such
regularization step is able to significantly improve the clustering
accuracy in high-dimensional clustering problems.

Another line of research focuses on simultaneous clustering and feature
selection. Some approaches are model-based clustering methods, such
as \citet{Raftery.Dean:2006}, \citet{Pan.Shen:2007}, \citet{Wang.Zhu:2008}, \citet{Xie.ea:2008}, and \citet{Guo.ea:2010}. In contrast,
some approaches are model-free, such as \citet{Witten.Tibshirani:2010}, \citet{Sun2012}, and \citet{Wang.ea:2013}. One common building block of
these sparse clustering approaches is the usage of a lasso-type penalty
for feature selection. For example, \citet{Witten.Tibshirani:2010}
developed a unified framework for feature selection in clustering using the
lasso penalty \citep{tibshirani:1996}. \citet{Sun2012} proposed a sparse k-means using the group-lasso penalty \citep{Yuan2006}.
We refer readers to \citet{Alelyani.ea:2013} for a thorough overview. In spite of their good numeric performance, these
sparse clustering procedures still suffer from instabilities due to
the non-convex optimization formulations. To overcome it, our sparse
convex clustering solves a convex optimization problem and ensures
a unique global solution.

\subsection{Paper Organization}

The rest of the manuscript is organized as follows. Section \ref{sec:main}
introduces the sparse convex clustering as well as its two efficient algorithms. Section \ref{sec:thm}
studies its statistical properties and Section \ref{sec:practical}
discusses some practical issues in the proposed implementations.
Section \ref{sec:experiment} evaluates the superior numeric performance
of the proposed methods through extensive simulations and a real data
application. Section \ref{sec:summary} finishes this paper with a brief summary. Technical details are provided in Appendix or online supplementary.

\section{Sparse Convex Clustering}

\label{sec:main}

This section presents the main results. We propose our new method, sparse convex clustering in Section \ref{sec:model}, and then we develop two efficient algorithms to implement the method in Section \ref{sec:algorithm}

\subsection{Model}

\label{sec:model}

To allow an adaptive penalization, we consider a modification of convex
clustering (\ref{eq:obj-cc}),

\begin{equation}
\min_{\mathbf{A}\in \mathbb{R}^{n\times p}}\frac{1}{2}\sum_{i=1}^{n}||X_{i\cdot}-A_{i\cdot}||_{2}^{2}+\gamma\sum_{i_{1}<i_{2}}w_{i_{1},i_{2}}||A_{i_{1}\cdot}-A_{i_{2}\cdot}||_{q},\label{eq:obj-cc-w}
\end{equation}
where the weight $w_{i_{1},i_{2}}\ge0$. \citet{Hocking2011} considered a pairwise
affinity weight $w_{i_{1},i_{2}}=\exp(-\phi\|X_{i_{1}\cdot}-X_{i_{2}\cdot}\|_{2}^{2})$
and \citet{Chi2015} suggested $w_{i_{1},i_{2}}=\iota_{i_{1},i_{2}}^{m}\exp(-\phi\|X_{i_{1}\cdot}-X_{i_{2}\cdot}\|_{2}^{2})$,
where $\iota_{i_{1},i_{2}}^{m}$ is 1 if observation $i_{2}$ is among
$i_{1}$'s $m$ nearest neighbors or vice verse, and 0 otherwise.

To introduce a reformulation of (\ref{eq:obj-cc-w}), we write the
data matrix $\mathbf{X}$ in feature-level as column vector $\mathbf{X}=(\mathbf{x}_{1},\cdots,\mathbf{x}_{p})$,
where $\mathbf{x}_{j}=(X_{1j},\cdots,X_{nj})^{\trans}$, $j=1,\ldots,p$
and denote $\mathbf{A}$ in feature-level as column vector $\mathbf{A}=(\mathbf{a}_{1},\cdots,\mathbf{a}_{p})$.
Without loss of generality, we assume the feature vectors are centered,
i.e., $\sum_{i=1}^{n}X_{ij}=0$ for each $j=1,\ldots, p$. Simple algebra implies that (\ref{eq:obj-cc-w})
can be reformulated as
\begin{equation} \label{eq:obj-cc-w-new}
\min_{\mathbf{A}\in \mathbb{R}^{n\times p}}\frac{1}{2}\sum_{j=1}^{p}||\mathbf{x}_{j}-\mathbf{a}_{j}||_{2}^{2}+\gamma\sum_{l\in{\cal E}}w_{l}||A_{i_{1}\cdot}-A_{i_{2}\cdot}||_{q},
\end{equation}
where ${\cal E}=\{l=(i_{1},i_{2}):1\leq i_{1}<i_{2}\leq n\}$.

For a given $\gamma$, let $\widehat{\mathbf{A}}=(\widehat{A}_{1\cdot},\cdots,\widehat{A}_{n.})^{\trans}=(\widehat{\mathbf{a}}_{1},\ldots,\widehat{\mathbf{a}}_{p})$
be the solution to (\ref{eq:obj-cc-w-new}). The clustering structure
is implied by the observation-level estimates, $\widehat{A}_{i\cdot}$,
$i=1,\ldots,n$; that is, if $\widehat{A}_{i_{1}\cdot}=\widehat{A}_{i_{2}\cdot}$,
then observations $i_{1}$ and $i_{2}$ are estimated to belong to
the same cluster. The feature importance is implied by the feature-level
estimates, $\widehat{\mathbf{a}}_{j}$, $j=1,\cdots,p$; that is, if the
components of a feature-level estimate $\widehat{\mathbf{a}}_{j}$ are
identical, then the corresponding feature $j$ is not informative
for clustering. Remind that the feature vectors are centered, then
feature $j$ is not informative if and only if $\|\widehat{\mathbf{a}}_{j}\|_{2}^{2}=\sum_{i=1}^{n}\widehat{A}_{ij}^{2}=0$.

In high-dimensional clustering, it is desired to have a sparse solution
$\wh{\bfA}$ with some of its column vectors being exact $\bfzero$'s.
Motivated by the importance of excluding non-informative features, we propose a new sparse convex clustering by incorporating
an adaptive group-lasso penalty \citep{Yuan2006,Wang.Leng:2008} into
the convex clustering objective function (\ref{eq:obj-cc-w-new}).
In particular, sparse convex clustering solves

\begin{equation}\label{eq:obj_constraint}
\min_{\mathbf{A}\in \mathbb{R}^{n\times p}}\frac{1}{2}\sum_{j=1}^{p}||\mathbf{x}_{j}-\mathbf{a}_{j}||_{2}^{2}+\gamma_{1}\sum_{l\in{\cal E}}w_{l}||A_{i_{1}\cdot}-A_{i_{2}\cdot}||_{q}+\gamma_{2}\sum_{j=1}^{p}u_{j}||\mathbf{a}_{j}||_{2},
\end{equation}
where tuning parameter $\gamma_{1}$ controls the cluster size and
tuning parameter $\gamma_{2}$ controls the number of informative
features. In the group-lasso penalty, the weight $u_{j}$ plays
an important role to adaptively penalize the features. Detailed discussions on practical choices of tuning parameters and weights can be found in Section \ref{sec:practical}.

\begin{remark}
In the objective function (\ref{eq:obj_constraint}) of sparse convex clustering, the second group-lasso-type penalty enforces the global sparsity condition; that is, the elements of each column vector $\bfa_j$ would be all zero or all nonzero. Such penalty is considered for the feature selection purpose. This global sparsity condition can be relaxed in two directions. First, we can replace the second penalty, $\gamma_2\sum_{j=1}^{p}u_{j}||\mathbf{a}_{j}||_{2}$, by a lasso type of penalty, $\gamma_2\sum_{j=1}^{p}||\mathbf{a}_{j}||_{1}$. Second, we can also add another penalty, $\sum_{j=1}^{p}||\mathbf{a}_{j}||_{1}$, to the objective function \ref{eq:obj_constraint}. This results in a new penalty, $\gamma_2\sum_{j=1}^{p}u_{j}||\mathbf{a}_{j}||_{2}+\gamma_3\sum_{j=1}^{p}||\mathbf{a}_{j}||_{1}$, which is the so-called sparse-group-lasso penalty \citep{Friedman2010}.
\end{remark}

\subsection{Algorithms}

\label{sec:algorithm}

This subsection discusses two efficient optimization approaches to
solve the sparse convex clustering by adopting a similar computational
strategy used in \citet{Chi2015}. Our two approaches are based on
the alternating direction method of multipliers (ADMM) algorithm \citep{Boyd2011,Gabay1976,Glowinski1975}
and the alternating minimization algorithm (AMA) \citep{Tseng1991},
and are referred as sparse ADMM (S-ADMM) and sparse AMA (S-AMA), respectively.

To implement the S-ADMM and S-AMA algorithms, we rewrite the
convex clustering problem in formula (\ref{eq:obj_constraint}) as
\begin{eqnarray*}
\min_{\bfA\in \mathbb{R}^{n\times p}} && \frac{1}{2}\sum_{j=1}^{p}||\mathbf{x}_{j}-\mathbf{a}_{j}||_{2}^{2}+\gamma_{1}\sum_{l\in{\cal E}}w_{l}||\mathbf{v}_{l}||_{q}+\gamma_{2}\sum_{j=1}^{p}u_{j}||\mathbf{a}_{j}||_{2}, \\
{\rm s.t.} && A_{i_1\cdot}-A_{i_2\cdot}-\bfv_l=\bfzero.
\end{eqnarray*}
This is equivalent to minimize the following augmented Lagrangian
function,
\begin{eqnarray*}
\mathcal{L}_{\nu}(\mathbf{A},\mathbf{V},\mathbf{\Lambda}) & = & \frac{1}{2}\sum_{j=1}^{p}\|\mathbf{x}_{j}-\mathbf{a}_{j}\|_{2}^{2}+\gamma_{1}\sum_{l\in{\cal E}}w_{l}\|\mathbf{v}_{l}\|_{q}+\gamma_{2}\sum_{j=1}^{p}u_{i}\|\mathbf{a}_{j}\|_{2}\\
 &  & +\sum_{l\in\mathbf{{\cal E}}}\langle\mathbf{\lambda}_{l},\mathbf{v}_{l}-A_{i_{1}\cdot}+A_{i_{2}\cdot}\rangle+\frac{\nu}{2}\sum_{l\in{\cal E}}\|\mathbf{v}_{l}-A_{i_{1}\cdot}+A_{i_{2}\cdot}\|_{2}^{2},
\end{eqnarray*}
where $\nu$ is a small constant, $\mathbf{V}=(\mathbf{v}_{1},\ldots,\mathbf{v}_{|{\cal E}|})$,
and $\mathbf{\Lambda}=(\mathbf{\lambda}_{1},\ldots,\mathbf{\lambda}_{|{\cal E}|})$.
Compared with the original algorithms proposed in \citet{Chi2015},
it becomes challenging to deal with the feature-level and observation-level
vectors in the new objective function simultaneously.

\subsubsection{S-ADMM}

S-ADMM minimizes the augmented Lagrangian problem by alternatively
solving one block of variables at a time. In particular, S-ADMM solves
\begin{eqnarray}
\mathbf{A}^{m+1} & = & \argmin_{\mathbf{A}}{\cal L}_{\nu}(\mathbf{A},\mathbf{V}^{m},\mathbf{\Lambda}^{m}),\nonumber \\
\mathbf{V}^{m+1} & = & \argmin_{\mathbf{V}}{\cal L}_{\nu}(\mathbf{A}^{m+1},\mathbf{V},\mathbf{\Lambda}^{m}),\\
\blambda_{l}^{m+1} & = & \blambda_{l}^{m}+\nu(\bfv_{l}^{m+1}-A_{i_{1}\cdot}^{m+1}+A_{i_{2}\cdot}^{m+1}), \ l \in \calE.\nonumber
\end{eqnarray}

Next we discuss the detailed updating implementations for $\mathbf{A},\mathbf{V}$
and $\mathbf{\Lambda}$ in three steps. A summary of the S-ADMM algorithm
is shown in Algorithm \ref{algADMM}.

\textbf{Step 1: update $\mathbf{A}$.} Denote $\wt{\bfv}_{l}=\bfv_{l}+\frac{1}{\nu}\blambda_{l}$.
Updating $\bfA$ is equivalent to minimizing
\begin{eqnarray}
f(\bfA)=\frac{1}{2}\sum_{j=1}^{p}\|\bfx_{j}-\bfa_{j}\|_{2}^{2}+\frac{\nu}{2}\sum_{l\in\calE}\|\wt{\bfv}_{l}-A_{i_{1}\cdot}+A_{i_{2}\cdot}\|_{2}^{2}+\gamma_{2}\sum_{j=1}^{p}u_{j}\|\bfa_{j}\|_{2}.\label{eqn:update_A}
\end{eqnarray}

This optimization problem is challenging because the objective function involves both rows and columns
of the matrix $\bfA$. To tackle this difficulty, the following key lemma associates $(\ref{eqn:update_A})$
with a group-lasso regression problem which can be efficiently
solved via standard packages.

\begin{lemma} \label{thm:equivalent}
Let $\bfI_{n}$ be an $n\times n$ identity
matrix, $\bfone_{n}$ be an $n$-dimensional
vector with each component being 1, and $\bfe_{i}$ be an $n$-dimensional
vector with each component being 0 but its $i$-th component being
1. Define $\bfN^{-1}=(1+n\nu)^{-1/2}[\bfI_{n}+n^{-1}(\sqrt{1+n\nu}-1)\bfone_{n}\bfone_{n}\trans]$
and denote $\bfy_{j}=\bfN^{-1}[\bfx_{j}+\nu\sum_{l\in\calE}\wt{v}_{jl}(\bfe_{i_{1}}-\bfe_{i_{2}})]$
with $\wt{v}_{jl}$ the $j$-th element of $\wt{\bfv}_{l}$. Then,
minimizing $(\ref{eqn:update_A})$ is equivalent to
\begin{eqnarray*}
\min_{\bfa_{j}}\frac{1}{2}\|\bfy_{j}-\bfN\bfa_{j}\|_{2}^{2}+\gamma_{2}u_{j}\|\bfa_{j}\|_{2}, \textrm{~for~each~} j=1,\ldots,p.
\end{eqnarray*}
\end{lemma}

The proof of Lemma \ref{thm:equivalent} is discussed in Appendix.
The key ingredient in the proof is a newly established property of
a permutation matrix, i.e., Proposition \ref{prop1}. Based on this
property, we are able to solve the minimization of $f(\bfA)$ by $p$
separate sub-optimization problems. This together with the property
of group-lasso penalty leads to desirable results. Recall that we require the feature vectors are centered, so we center corresponding estimates during each iteration.

\textbf{Step 2: update $\bfV$.} For any $\sigma>0$ and norm $\Omega(\cdot)$,
we define a proximal map,
\begin{eqnarray*}
\textrm{prox}_{\sigma\Omega}(\bfu)=\argmin_{\bfv}\left[\sigma\Omega(\bfv)+\frac{1}{2}\|\bfu-\bfv\|_{2}^{2}\right].
\end{eqnarray*}
In S-ADMM, $\Omega(\cdot)$ is a $q$-norm $\|\cdot\|_{q}$ with $q=1,2$,
or $\infty$, and $\sigma=\gamma_{1}w_{l}/\nu$. We refer the readers to Table 1 of \citet{Chi2015} for the explicit formulations
of the proximal map of $q$-norm for $q=1,2$ and $\infty$. Because vectors ${\bfv}_{l}$ are
separable, they can be solved via proximal maps, that is
\begin{eqnarray*}
\bfv_{l} & = & \argmin_{\bfv_{l}}\frac{1}{2}\|\bfv_{l}-(A_{i_{1}\cdot}-A_{i_{2}\cdot}-\nu^{-1}\blambda_{l})\|_{2}^{2}+\frac{\gamma_{1}w_{l}}{\nu}\|\bfv_{l}\|_{q}\\
 & = & \textrm{prox}_{\sigma_{l}\|\cdot\|_{q}}(A_{i_{1}\cdot}-A_{i_{2}\cdot}-\nu^{-1}\blambda_{l}).
\end{eqnarray*}

\textbf{Step 3: update $\bLambda$.} Finally, $\blambda_{l}$ can
be updated by $\blambda_{l}=\blambda_{l}+\nu(\bfv_{l}-A_{i_{1}\cdot}+A_{i_{2}\cdot})$.

\begin{algorithm}[!htb]
\protect\caption{\quad{}S-ADMM \label{algADMM}}

\begin{enumerate}
\item Initialize $\mathbf{V}^{0}$ and $\mathbf{\Lambda}^{0}$. For $m=1,2,\ldots$
\item For $j=1,\ldots,p$, do
\begin{eqnarray*}
\wt{\bfv}_{l}^{m-1} & = & \bfv_{l}^{m-1}+\frac{1}{\nu}\blambda_{l}^{m-1}, l\in\calE \\
\bfy_{j}^{m-1} & = & \bfN^{-1}\left(\bfx_{j}+\nu\sum_{l\in\calE}\wt{v}_{lj}^{m-1}(\bfe_{i_{1}}-\bfe_{i_{2}})\right),\\
\bfa_{j}^{m} & = & \argmin_{\bfa_{j}}\frac{1}{2}\|\bfy_{j}^{m-1}-\bfN\bfa_{j}\|_{2}^{2}+\gamma_{2}u_{j}\|\bfa_{j}\|_{2}, \\
\bfa_{j}^{m} &=& \bfa_{j}^{m} - \widebar\bfa_{j}^{m}\bfone_n, \ {\rm where} \ \widebar\bfa_{j}^{m}=\bfone_n^T \bfa_{j}^{m}/n.
\end{eqnarray*}

\item For $l\in\calE$, do
\begin{eqnarray*}
\bfv_{l}^{m}=\textrm{prox}_{\sigma_{l}\|\cdot\|_{q}}(A_{i_{1}\cdot}^{m}-A_{i_{2}\cdot}^{m}-\nu^{-1}\blambda_{l}^{m-1}).
\end{eqnarray*}

\item For $l\in\calE$, do
\begin{eqnarray*}
\blambda_{l}^{m}=\blambda_{l}^{m-1}+\nu(\bfv_{l}^{m}-A_{i_{1}\cdot}^{m}+A_{i_{2}\cdot}^{m}).
\end{eqnarray*}

\item Repeat Steps 2-4 until convergence. \end{enumerate}
\end{algorithm}

\subsubsection{S-AMA}

To increase the computational efficiency, we introduce another algorithm S-AMA for implementing sparse convex clustering. S-AMA is different
from S-ADMM in the update of $\bfA$. In particular, S-AMA solves
$\bfA$ by treating $\nu=0$, i.e., $\bfA^{m+1}=\argmin_{\bfA}\calL_{0}(\bfA,\bfV^{m},\bLambda^{m})$.
When $\nu=0$, we have $\mathbf{N}=\mathbf{I}_{n}$ and $\bfy_{j}=\bfx_{j}$.
According to Lemma \ref{thm:equivalent}, updating $\bfA$ requires to solve $p$
group-lasso problems:
\begin{equation}\label{eq:g1}
\min_{\bfa_{j}}\frac{1}{2}\|\mathbf{x}_{j}-\mathbf{a}_{j}\|_{2}^{2}+\gamma_{2}u_{j}\|\mathbf{a}_{j}\|_{2}, j=1,\ldots,p.
\end{equation}
By Karush-Kuhn-Tucker (KKT) conditions of the group lasso problem \citep{Yuan2006}, the solution
to (\ref{eq:g1}) has a closed form as
\begin{eqnarray*}
\wh\bfa_{j}=\left(1-\frac{\gamma_{2}u_{j}}{\|\bfz_{j}\|_{2}}\right)_{+}\bfz_{j},
\end{eqnarray*}
where $\bfz_{j}=\bfx_{j}+\sum_{l\in\calE}\lambda_{jl}(\bfe_{i_{1}}-\bfe_{i_{2}})$
and $(z)_{+}=\max\{0,z\}$. See the detailed derivations in online Supplementary. Still, we center $\wh\bfa_j$ for each $j$.
The above formula significantly reduces the computational cost by solving
$p$ group-lasso problem analytically in each iteration. Note that
the above update of $\bfA$ is independent of $\bfV$, which indicates
that S-AMA algorithm does not need to compute the update of $\bfV$.
Therefore S-AMA is much more efficient than S-ADMM algorithm.

Next, we discuss the update of $\bLambda$. Define $\calP_{B}(\bfz)$
as a projection onto $B=\{\bfy:\|\bfy\|_{\dag}\leq1\}$ of the norm
$\|\cdot\|_{\dag}$, where $\|\cdot\|_{\dag}$ is the dual norm of
$\|\cdot\|_{q}$, which defines the fusion penalty. We show in
online Supplementary that the update of $\bLambda$ reduces to $\blambda_{l}^{m}= \calP_{C_{l}}[\blambda_{l}^{m-1}-\nu(A_{i_{1}\cdot}^{m}-A_{i_{2}\cdot}^{m})]$
with $C_{l}=\{\blambda_{l}:\|\blambda_{l}\|_{\dag}\leq\gamma_{1}w_{l}\}$.
The S-AMA algorithm is summarized in Algorithm \ref{algAMA}.

\begin{algorithm}[!htb]
\protect\caption{\quad{}S-AMA \label{algAMA}}

\begin{enumerate}
\item Initialize $\mathbf{\Lambda}{}^{0}$. For $m=1,2,\ldots$
\item For $j=1,\ldots,p$, do
\begin{eqnarray*}
\bfz_{j}^{m} & = & \bfx_{j}+\sum_{l \in\calE}\lambda_{lj}^{m-1}(\bfe_{i_{1}}-\bfe_{i_{2}}),\\
\bfa_{j}^{m} & = & \left(1-\frac{\gamma_{2}u_{i}}{\|\bfz_{i}^{m}\|_{2}}\right)_{+}\bfz_{j}^{m},\\
\bfa_{j}^{m} &=& \bfa_{j}^{m} - \widebar\bfa_{j}^{m}\bfone_n, \ {\rm where} \ \widebar\bfa_{j}^{m}=\bfone_n^T \bfa_{j}^{m}/n.
\end{eqnarray*}

\item For $l\in\calE$, do
\[
\blambda_{l}^{m}=\calP_{C_{l}}[\blambda_{l}^{m-1}-\nu(A_{i_{1}\cdot}^{m}-A_{i_{2}\cdot}^{m})],
\]
where $C_{l}=\{\blambda_{l}:\|\blambda_{l}\|_{\dag}\leq\gamma_{1}w_{l}\}$.
\item Repeat Steps 2-3 until convergence. \end{enumerate}
\end{algorithm}

\subsubsection{Algorithmic Convergence}

This subsection discusses the convergence of the proposed S-ADMM and
S-AMA algorithms. \citet{Chi2015} and the references therein provided sufficient conditions for the convergence
of the following general optimization problem,
\begin{eqnarray}
\min_{\xi,\zeta}\ f(\xi)+g(\zeta),\textrm{\ \ s.t.\ \ }A\xi+B\zeta=c.\label{general_obj}
\end{eqnarray}
They verified that the ADMM and AMA algorithms for
convex clustering, as two special cases of \eqref{general_obj}, satisfied
the sufficient conditions under which the convergence was guaranteed.

The convergence of our S-ADMM and S-AMA algorithms follows similar
arguments. Note that the only difference between the objective function
in \eqref{eq:obj_constraint} and its counterpart in \citet{Chi2015}
is a convex penalty term $\gamma_{2}\sum_{j=1}^{p}u_{j}\|\mathbf{a}_{j}\|_{2}$.
Define the summation of the first and third terms of the objective
function in \eqref{eq:obj_constraint} as $f(\cdot)$, and the second
term as $g(\cdot)$. This indicates that the optimization problem
\eqref{eq:obj_constraint} is a special case of \eqref{general_obj}.
Simple algebra implies that $f(\cdot)$ is strongly convex. According
to \citet{Chi2015}, one can show that, under mild regularization conditions, the convergence
of S-ADMM is guaranteed for any $\nu>0$, and the convergence
of S-AMA algorithm is guaranteed provided that positive
constant $\nu$ is not too large.

\subsubsection{Computational Consideration}

Step 2 in both Algorithms \ref{algADMM} and \ref{algAMA} involves $p$ sub-optimization problems. Therefore, S-ADMM and S-AMA merit from the distributed optimization, and they can handle large-scale problems efficiently. To be specific, Step 2 can be distributed to different processors to obtain estimates of $\bfa_j$'s which are then gathered to update $\bfA$. In addition, Steps 3-4 in Algorithm \ref{algADMM} or Step 3 in Algorithm \ref{algAMA} can also be distributed to different processors to obtain fast updates.

It is worth pointing out that the computation of S-AMA is comparable to AMA in \cite{Chi2015}, while S-ADMM is computationally more expensive than ADMM in \cite{Chi2015} and S-AMA. This is because Step 2 in S-ADMM does not have a closed-form formula and it requires solving $p$ group-lasso problems assisted by iterations. Furthermore, Step 3 in S-AMA only requires updates for $\blambda_l$ for those $l$ such that $w_l > 0, l \in \calE$. With a suitable selection of $w_l$ discussed in Section \ref{sec:weights}, the size of working set of $l$ can be dramatically reduced from $n(n-1)/2$ to a much smaller number. Our limited experience in numerical studies also confirms the superiority of S-AMA over S-ADMM in terms of the computational cost.

We have developed an $\textsf{R}$ package ``\textit{scvxclustr}" to facilitate the implementation of proposed methods. Table \ref{tb_time} compares the computational time in seconds of our package with \cite{Chi2015}'s method via ``\textit{cvxclustr}" for the setting with 4 clusters described in Section \ref{sec:simulation} with given tuning parameters. The computer is equipped with a CPU i3-4170 (3.70GHz) and 8G memory.

\begin{table}[!htb]
\centering \protect\caption{Timing comparison under the 4 cluster setting described in Section \ref{sec:simulation} with given tuning parameters under various combinations of $n$ and $p$. The time is in second.}
\label{tb_time} \vspace{6pt}
 \centering  \small
\begin{tabular}{L{1.7cm}R{0.7cm}R{0.7cm}R{0.7cm}p{0.01cm}R{0.7cm}R{0.7cm}R{0.7cm}p{0.01cm}R{0.8cm}R{0.8cm}R{0.8cm}R{0.01cm}R{0.7cm}R{0.7cm}R{0.7cm}}
\hline
$n$  & \multicolumn{3}{c}{60} & \  & \multicolumn{3}{c}{120} & \  & \multicolumn{3}{c}{300} & \  & \multicolumn{3}{c}{400}\tabularnewline
\hline
$p$ & 150  & 300  & 500  & \  & 150  & 300  & 500  & \  & 150  & 300  & 500 & \ & 150  & 300  & 500 \tabularnewline
\hline
\hline
AMA	& 0.6 &	3 &	5 & \ &	3 &	7 &	13 & \ &	11 &	23 &	37 & \ &	15 &	32 &	51 \tabularnewline
ADMM &	6 &	17 &	24 & \ &	19 &	71 &	82 & \ &	157 &	375 &	548 & \ &	329 &	1482	& 1796 \tabularnewline
S-AMA &	0.1 &	0.6 &	2 & \ &	0.8 &	2 &	6 & \ &	3 & 10 &	25 & \ &	6 &	17 &	34 \tabularnewline
S-ADMM	& 16	& 47 &	60 & \ &	60 &	152 &	215 & \ &	593 &	2051 &	2781 & \ &	1252 &	3639 &	7925 \tabularnewline
\hline
\end{tabular}
\end{table}

\section{Theoretical Properties} \label{sec:thm}

In this section, we study statistical properties of the proposed sparse convex clustering. In particular, we provide finite sample bounds for prediction error of the proposed sparse convex clustering estimator and establish its variable selection consistency. In Section S.3 of online Supplementary, we also develop unbiased estimators for the degrees of freedoms of sparse convex clustering.

Assume $\bfx=\bfa_0 + \bvarepsilon$, where $\bvarepsilon \in \real^{np}$ is a vector of
independent sub-Gaussian noise terms with mean zero and variance $\sigma^2$, and $\bfa_0={\rm vec}(\bfA_0)=(\bfa_{01}\trans,\ldots, \bfa_{0p}\trans)\trans$
is a $np$-dimensional mean vector. Without loss of generality, we assume that only the first $p_0 < p$
features are informative, i.e., $\| \bfa_{0j} \|_2 \neq 0$ for $j \leq p_0$ and $\| \bfa_{0j} \|_2 = 0$ for $j > p_0$.
The informative feature set is denoted as $\calA=\{1,\ldots,p_0 \}$ and the noninformative feature set
is $\calA^c = \{p_0+1,\ldots,p \}$. For simplicity, we consider the case with $w_l=1$.

Our sparse convex clustering in (\ref{eq:obj_constraint}) can be reformulated as the following problem:
\begin{eqnarray} \label{eq:obj_bound}
\wh{\bfa} = \argmin_{\bfa \in \real^{np}} \frac{1}{2} \| \bfx - \bfa \|_2^2 + \gamma_1 \sum_{l \in \calE} \|\bfC_l \bfa \|_q
+ \gamma_2 \sum_{j=1}^{p} u_j \| \bfa_j \|_2,
\end{eqnarray}
where $\bfC_l=\bfI_p \otimes (\bfe_{i_1}-\bfe_{i_2})\trans$ and hence $\bfC_l \bfa = A_{i_1 \cdot}- A_{i_2 \cdot}$. Define $\bfC = ( \bfC_1 \trans, \ldots, \bfC_{|\calE|}\trans )\trans$ and denote $\bfu=(u_1,\ldots,u_p)\trans$.

The following two theorems provide the finite sample bounds for prediction error of our sparse convex clustering estimator with parameter $q \in \{1,2\}$, respectively.

\begin{theorem} \label{thm:bound1}
Let $\wh\bfa$ be the estimate of \eqref{eq:obj_bound} with $q=1$. If $\gamma_1 > 4 \sigma \sqrt{\frac{\log (p \cdot \binom{n}{2})}{n}}$, then
\begin{eqnarray*}
\frac{1-\gamma_2}{2np}  \| \wh\bfa - \bfa_0 \|_2^2  \leq
\frac{3\gamma_1}{2np} \| \bfC \bfa_0 \|_1 + \frac{\gamma_2 \| \bfu \|_2^2 }{2np} +\sigma^2 \left[
\frac{1}{n} + \sqrt{\frac{\log (np)}{n^2 p}} \right]  + \frac{1}{np}
\end{eqnarray*}
holds with probability at least $1-c_3$, where
\begin{eqnarray*}
c_3=\frac{2}{p \cdot \binom{n}{2}} +\exp\{ - \min (c_1 \log (np),c_2 \sqrt{p \log (np)} )\} + 2 \exp \big( -np/(2 \sigma^2 \gamma_2^2  \| \bfu \|_1^2) \big),
\end{eqnarray*}
for some positive constants $c_1$ and $c_2$ defined in Lemma \ref{lemma:HW71}.
\end{theorem}

\begin{theorem} \label{thm:bound2}
Let $\wh\bfa$ be the estimate of \eqref{eq:obj_bound} with $q=2$. If $\gamma_1 > 4 \sigma \sqrt{\frac{\log (p \cdot \binom{n}{2})}{n}}$, then
\begin{eqnarray*}
\frac{1-\gamma_2}{2np}  \| \wh\bfa - \bfa_0 \|_2^2  \leq
\frac{3\gamma_1}{2np} \sum_{l\in\calE} \| \bfC_l \bfa_0 \|_2+ \frac{\gamma_2 \| \bfu \|_2^2 }{2np} +\sigma^2 \left[
\frac{1}{n} + \sqrt{\frac{\log (np)}{n^2 p}} \right]  + \frac{1}{np}
\end{eqnarray*}
holds with probability at least $1-c_3$, where $c_3$ is defined in Theorem \ref{thm:bound1}.
\end{theorem}

\begin{remark}
Based on the sparsity assumption of features, we know true underlying clusters differ only with respect to the first $p_0$ features. Thus, $\| \bfC_l \bfa_0 \|_1=O(1)$ and $\| \bfC_l \bfa_0 \|_2=O(1)$, $\forall l \in \calE$. Note that $| \calE| = \binom{n}{2} $, and hence $\|\bfC \bfa_0 \|_1 = O(n^2)$ and $\sum_{l \in \calE} \| \bfC_l \bfa_0 \|_2 =O(n^2)$. In order to obtain a general prediction consistency based on Theorems \ref{thm:bound1}-\ref{thm:bound2}, we need $\gamma_2 \to 0, c_3 \to 0$ and the first two terms on right-hand side of the inequalities in Theorems \ref{thm:bound1}-\ref{thm:bound2} vanish. Assume $\gamma_1 \| \bfC \bfa_0 \|_1/ (2np)=o(1)$, and then we have $\sqrt{ n \log (p \cdot \binom{n}{2})/p^2}=o(1)$. Additionally, $c_3 \to 0$ and $\gamma_2 \| \bfu \|_2^2 / (2np)$ are equivalent to requiring $\gamma_2 \| \bfu \|_1^2 /(np) \to 0$ by noting that $\| \bfu \|_2 \leq \| \bfu \|_1$.

In particular, we discuss following two choices for the weight $u_j$, which lead to prediction consistency. First, we assume a non-adaptive weight, i.e., $u_j\equiv1$. It is easy to see $\| \bfu \|_1=p$. Additionally, we require $\gamma_2 \to 0$ and $\gamma_2 p n^{-1} \to 0$. Second, more generally, we assume $ \| \bfu \|_1$ is bounded above by $p^{\zeta}$, where $\zeta$ is a constant, and require $\gamma_2 \to 0$, $\gamma_2 p^{2\zeta-1} n^{-1} \to 0$. Therefore, in either case $\gamma_2 \| \bfu \|_1^2 /(np) \to 0$ and then $\wh\bfa$ is prediction consistent with $q=1$ or $q=2$.
\end{remark}

Next, we establish the asymptotic selection consistency of the proposed sparse convex clustering, which is a desirable property in high-dimensional cluster analysis where many features are non-informative.

\begin{theorem} \label{thm:selection_consistency}
If $\gamma_1 > 4 \sigma \sqrt{\log (p \cdot \binom{n}{2})/n}$, $\gamma_1 \| \bfC \bfa_0 \|_1/ (2np)=o(1)$, $\gamma_2 \to 0$ and $\gamma_2 \| \bfu \|_1^2 /(np) \to 0$ as $n,p \to \infty$, then $P(\|\wh\bfa_j \|_2=0)\to 1$ for any $j \in \calA^c$, with the solution $\wh\bfa$ to \eqref{eq:obj_bound} with either $q=1$ or $q=2$.
\end{theorem}

\begin{remark}
Condition $\gamma_2 \| \bfu \|_1^2 /(np) \to 0$ generally implies that the adaptive weights cannot be too large. For example, uniform weights satisfy this condition. Note that Conditions $\gamma_1 > 4 \sigma \sqrt{\log (p \cdot \binom{n}{2})/n}$ and $\gamma_1 \| \bfC \bfa_0 \|_1/ (2np)=o(1)$ imply $\sqrt{ n \log (p \cdot \binom{n}{2})/p^2}=o(1)$. This is derived to ensure the estimation consistency in Theorems \ref{thm:bound1} and \ref{thm:bound2}, in particular, to ensure the first term on the right-hand side of Theorem \ref{thm:bound1} or \ref{thm:bound2} to converge to zero. This condition requires that $n = o(p^2)$ up to a $\log$-term, and hence is satisfied as long as $p$ diverges not too slow, which is typically true in high-dimensional scenarios. Note that in this case $p$ can still be smaller than $n$. Similar phenomenon has also been found in \citet{Tan2015}.
\end{remark}

Theorem \ref{thm:selection_consistency} establishes the asymptotic selection consistency in the sense that
the proposed sparse convex clustering can eliminate the non-informative variables in
the estimated cluster centers with probability tending to one.This variable selection consistency is illustrated in the motivation example shown in Figure \ref{heatmap}.

Proofs of Theorems \ref{thm:bound1}-\ref{thm:selection_consistency} are provided in online Supplementary.

\section{Practical Issues}

\label{sec:practical}

In Section \ref{sec:algorithm}, the S-ADMM and S-AMA algorithms rely
on the choice of weights and the tuning parameters $\gamma_{1}$ and
$\gamma_{2}$. In this section, we discuss how to choose these parameters
in practice.

\subsection{Selection of Weights}
\label{sec:weights}

This subsection introduces practical selections of the weights $w_{i_{1},i_{2}}$,
$(i_{1},i_{2})\in\calE$, in the fused-lasso penalty, and the factors
$u_{j}$, $j=1,\cdots,p$, in the adaptive group-lasso penalty.

Following \citet{Chi2015}, we choose weights by incorporating the
m-nearest-neighbors method with Gaussian kernel. To be specific, the
weight between the sample pair $(i_{1},i_{2})$ is set as $w_{i_{1},i_{2}}=\iota_{i_{1},i_{2}}^{m}\exp(-\phi\|X_{i_{1}\cdot}-X_{i_{2}\cdot}\|_{2}^{2})$,
where $\iota_{i_{1},i_{2}}^{m}$ equals 1 if observation $i_{2}$
is among observation $i_{1}$'s $m$ nearest neighbors or vice versa, and
0 otherwise. This choice of weights works well for a wide range of
$\phi$ when $m$ is small. In our numerical results, $m$ is fixed
at $5$ and $\phi$ is fixed at 0.5.

Next we consider the selection of factor $u_{j}$. As suggested
by \citet{zou:2006}, $u_{j}$ can be chosen as $1/\|\wh{\bfa}_{j}^{(0)}\|_{2}$,
where $\wh\bfa_{j}^{(0)}$ is the estimate of $\bfa_{j}$ in \eqref{eq:obj_constraint}
with $\gamma_{2}=0$. Such choice of factors penalizes less on informative
features and penalizes more on uninformative features, and hence leads
to improved clustering accuracy and variable selection performance
than its non-adaptive counterpart.

Finally, in order to ensure that the optimal tuning parameters $\gamma_{1}$
and $\gamma_{2}$ lie in relatively robust intervals regardless of
feature dimension and sample size, weights $w_{i_{1},i_{2}}$ and
factors $u_{j}$ are re-scaled to sum to $1/\sqrt{p}$ and $1/\sqrt{n}$,
respectively. Such re-scaling is only for convenience and does not
affect the final clustering path.

\subsection{Selection of Tuning Parameters}

\label{sec:tuning}

This subsection provides a selection method for tuning parameters
$\gamma_{1}$ and $\gamma_{2}$. Remind that $\gamma_{1}$ controls
the number of estimated clusters and $\gamma_{2}$ controls the number
of selected informative features.

We first illustrate via a toy example the effectiveness of tuning
parameter $\gamma_{2}$ on variable selection accuracy. In this
example, $60$ observations with $p=500$ features are generated from
4 clusters. Among all the features, only $20$ variables differ between
clusters. See detailed simulation setup in Section \ref{sec:simulation}.
By fixing $\wh\gamma_{1}=2.44$ and varying $\gamma_{2}$ from $e^{-5.0}$
to $e^{7.0}$, we plot the path of false negative rate (FNR) and the
path of false positive rate (FPR) of the final estimator. As shown
in Figure \ref{demo-graph}, when $\gamma_{2}$ is close to zero,
all features are included, and when $\gamma_{2}$ increases to some
ranges of intervals, all and only uninformative features are excluded,
i.e., perfect variable selection performance. This illustrates the
sensitivity of $\gamma_{2}$ to the variable selection performance
of the final estimator. In practice, we aim to estimate a suitable
$\gamma_{2}$ that leads to satisfactory variable selection.

\begin{figure}[!htb]
\protect\caption{Illustration of the effectiveness of $\gamma_{2}$ on variable selection
accuracy. The solid curve is the path of false negative rate (FNR),
and the dashed curve is the path of false positive rate (FPR).}
\centering \includegraphics[scale=0.5]{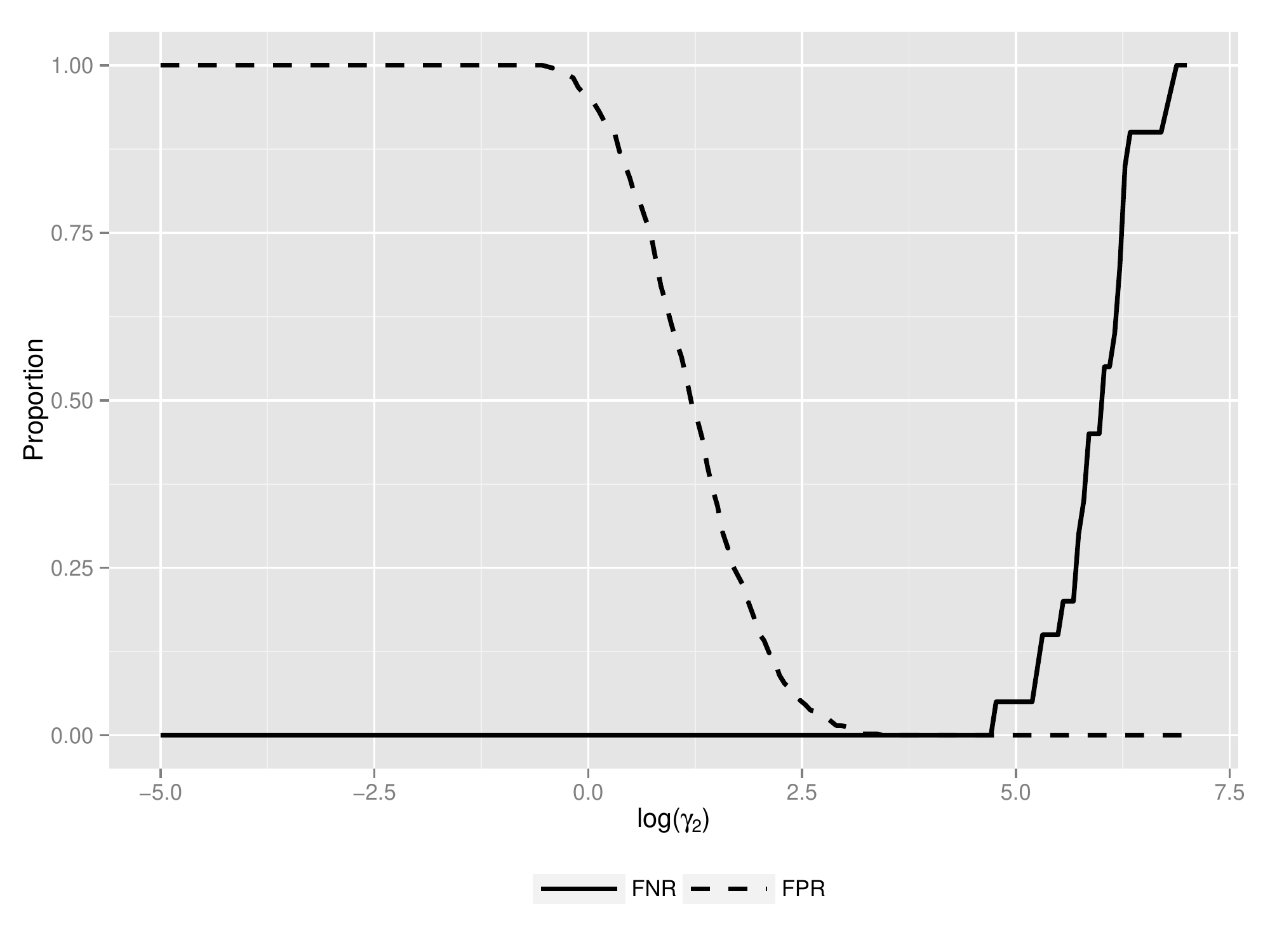}\label{demo-graph}
\end{figure}

In literature, \citet{Wang:2010} and \citet{Fang2012} proposed stability
selection to estimate the tuning parameters in clustering models.
The idea behind stability selection is that a good tuning parameter
should produce clustering results that are stable with respect to
a small perturbation to the training samples. Stability selection well suits
the model selection in cluster analysis because cluster labels are unavailable and the cross-validation method is not applicable in this case.

In this paper, we propose
to use stability selection in \citet{Fang2012} to tune both parameters
$\gamma_{1}$ and $\gamma_{2}$. To be specific, for any given $\gamma_{1}$
and $\gamma_{2}$, based on two sets of bootstrapped samples, two
clustering results can be produced via \eqref{eq:obj_constraint}, and
then the stability measurement \citep{Fang2012} can be computed to
measure the agreement between the two clustering results. In order
to enhance the robustness of the stability selection method, we repeat
this procedure $50$ times and then compute the averaged stability
value. Finally, the optimal parameter is selected as the one achieving
maximum stability. Our extensive numerical studies show that the selection of
important features is less sensitive to the clustering path, i.e., import features stand out in almost all clustering structures.
Thus, to speed up tuning process, stability path can be computed over
of a coarse grid of $\gamma_{1}$ and a fine grid of $\gamma_{2}$.

\section{Numerical Results}

\label{sec:experiment}

This section demonstrates the superior performance of our sparse convex
clustering in simulated examples in Section \ref{sec:simulation}
and a real application of hand movement clustering in Section \ref{sec:real}.

\subsection{Simulation Studies}

\label{sec:simulation}

In this subsection, simulations studies are conducted to evaluate
the performance of sparse convex clustering methods (S-ADMM and S-AMA). They
are compared to the k-means clustering and two convex clustering algorithms: ADMM and AMA \citep{Chi2015}.

First, we consider four spherical settings. Each simulated dataset consists
of $n=60$ observations with the number of clusters either $K=2$
or 4, and the number of features either $p=150$ or $500$. In each
setting, only the first 20 features are informative and remaining
features are non-informative. The samples $X_{i\cdot}\in\real^{p},i=1,\ldots,n$,
are generated as follows. Denote a $p$-dimensional multivariate normal distribution as $\textrm{MVN}_{p}$. For each $i$, a cluster label $Z_{i}$
is uniformly sampled from $\{1,\ldots,K\}$, and then the first 20
informative features are generated from $\textrm{MVN}_p(\bmu_{K}(Z_{i}),\bfI_{20})$,
where $\bmu_{K}(Z_{i})$ is defined as follows:
\begin{itemize}
\item If $K=2$, $\bmu_{2}(Z_{i})=\mu\bfone_{20}I(Z_{i}=1)-\mu\bfone_{20}I(Z_{i}=2)$;
\item If $K=4$, $\bmu_{4}(Z_{i})=(\mu\bfone_{10}\trans,-\mu\bfone_{10}\trans)\trans I(Z_{i}=1)+(-\mu\bfone_{10}\trans,-\mu\bfone_{10}\trans)\trans I(Z_{i}=2)+\ \ \ \ \ \ \ \ \ \ \ \ \ (-\mu\bfone_{10}\trans,\mu\bfone_{10}\trans)\trans I(Z_{i}=3)+(\mu\bfone_{10}\trans,\mu\bfone_{10}\trans)\trans I(Z_{i}=4)$,
\end{itemize}
where $\mu$ controls the
distance between cluster centers. Here a large $\mu$ indicates that
clusters are well-separated, whereas a small $\mu$ indicates that
clusters are overlapped. Finally, the rest $p-20$ noise features
are generated from $\calN(0,1)$.

Second, we consider a non-spherical setting with two half moons. Each simulated dataset consists of $n=100$ observations with $K=2$ clusters and $p=40$ features. Only the first two features are informative, and the rest 38 noisy features are generated from $\calN (0, 0.5)$. It is a relatively hard setting because the number of non-informative features are 19 times more than that of informative features. Figure \ref{twohalfmoons} shows one example of two interlocking half moons with the first two features. For a comparison purpose, we also apply the spectral clustering (SPECC, see \cite{ng2002spectral}) via an $\textsf{R}$ package ``\textit{kernlab}"  because SPECC can internally tackle non-spherical clusters.

\begin{figure}[!htb]
\protect\caption{The plot of the first two features for one example of two interlocking half moons.}
\centering \includegraphics[scale=0.45]{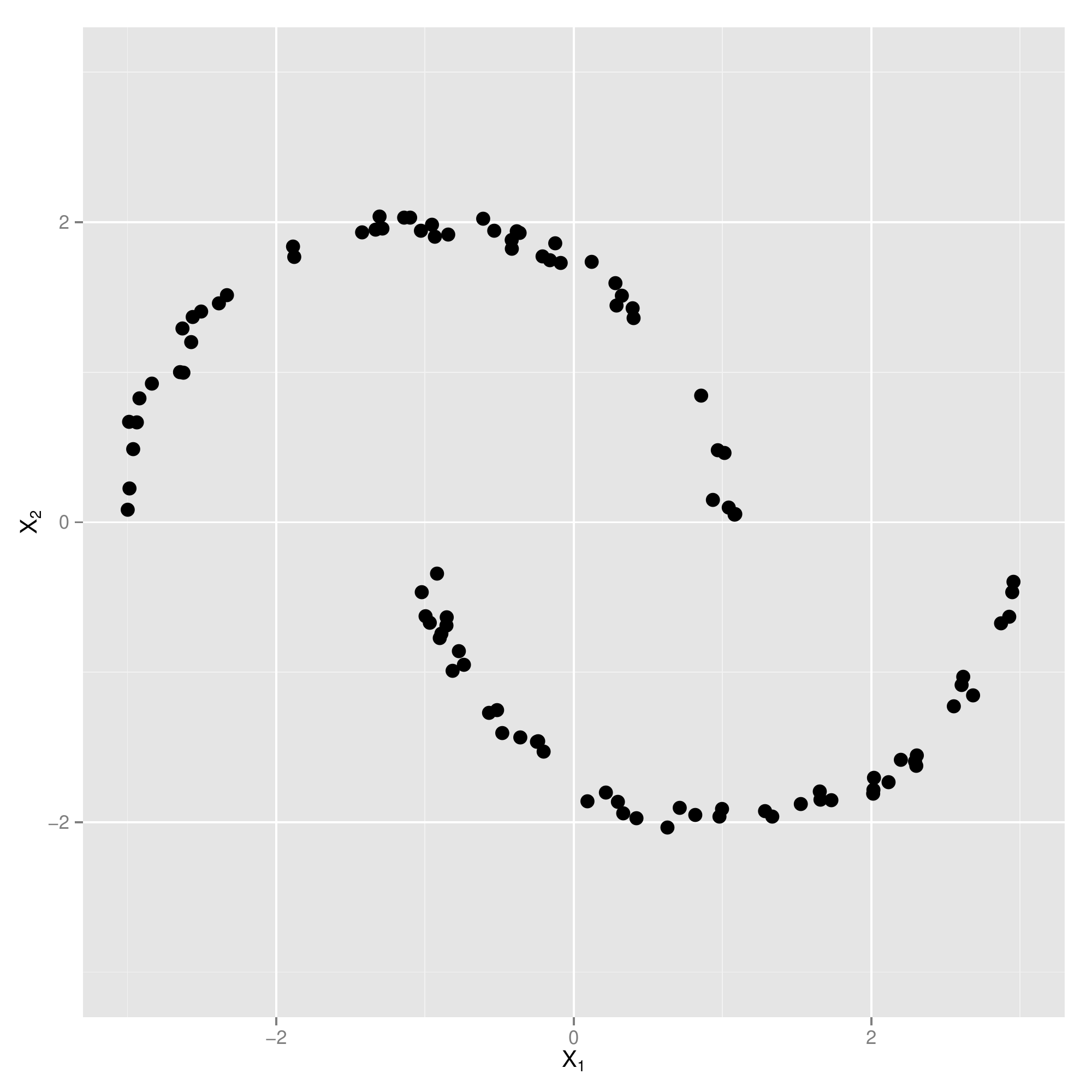} \label{twohalfmoons}
\end{figure}

In summary, five simulation settings are considered. Setting 1: $K=2,n=60, p=150$,
and $\mu=0.6$; Setting 2: $K=2,n=60,p=500$, and $\mu=0.7$; Setting 3:
$K=4,n=60,p=150$, and $\mu=0.9$; Setting 4: $K=4,n=60,p=500$, and $\mu=1.2$; Setting 5: two half moons with $K=2, n=100, p=40$.
For each setting, we run 200 repetitions.

The RAND index \citep{Rand1971} is used to measure the agreement
between the estimated clustering result and the underlying true clustering
assignment. The RAND index ranges between 0 and 1, and a higher value
indicates better performance. Note that the true cluster labels are known in
simulation studies, and thus it is feasible to know how well the candidate methods can
perform if they are tuned by maximizing the RAND index.
To ensure fair comparisons, for each
repetition, separate validation samples are generated and used to
select an optimal $k$ in k-means, an optimal $\gamma$ in ADMM and AMA, and optimal $\gamma_{1}$
and $\gamma_{2}$ in S-ADMM and S-AMA. To evaluate the performance
of variable selection, two measurements are reported: the false negative
rate (FNR) and the false positive rate (FPR). All the simulation results
are summarized in Table \ref{tb_simulation_1}. Due to its relatively
expensive computational costs, S-ADMM is not evaluated for Settings
2 and 4 where $p=500$.

In the first four simulation settings, the centers are spherical and
hence k-means always performs well in clustering accuracy, i.e., large
RAND index. The goals of these simulations are to justify that (1)
convex clustering does not perform well when the feature dimension
is high; (2) sparse convex clustering performs very well when the
feature dimension is high; and (3) sparse convex clustering selects
informative features with great clustering accuracy. The goal of Setting 5 is to show under the non-spherical setting, sparse convex clustering can still perform well, even better than the SPECC. All these claims
are justified by the results presented in Table \ref{tb_simulation_1}.

\begin{table}[!htb]
\centering \protect\caption{Empirical mean and standard deviation (SD) of the RAND index, false
positive rate (FPR), and false negative rate (FNR) based on 200 repetitions.
Setting 1: $K=2,n=60,p=150$, and $\mu=0.6$; Setting 2: $K=2,n=60,p=500$,
and $\mu=0.7$; Setting 3: $K=4,n=60,p=150$, and $\mu=0.9$; Setting 4:
$K=4,n=60,p=500$, and $\mu=1.2$; Setting 5: two half moons with $K=2, n=100, p=40$. The best RAND index in each scenario is shown
in bold. }

\label{tb_simulation_1} \vspace{6pt}
 \centering %
\begin{tabular}{p{2cm}p{2cm}p{1.2cm}p{1.2cm}p{0.1cm}p{1.2cm}p{1.2cm}p{0.1cm}p{1.2cm}p{1.2cm}}
\hline
\  & \  & \multicolumn{2}{c}{RAND} & \  & \multicolumn{2}{c}{FNR} & \  & \multicolumn{2}{c}{FPR}\tabularnewline
\hline
 & Algorithm  & mean  & SD  & \  & mean  & SD  & \  & mean  & SD \tabularnewline
\hline
\hline
Setting 1  & k-means  & 0.95  & 0.06  & \  & 0.00  & 0.00  & \  & 1.00  & 0.00 \tabularnewline
\  & ADMM  & 0.53  & 0.39  & \  & 0.00  & 0.00  & \  & 1.00  & 0.00 \tabularnewline
\  & AMA  & 0.66  & 0.40  & \  & 0.00  & 0.00  & \  & 1.00  & 0.00 \tabularnewline
\  & S-ADMM  & 0.82  & 0.24  & \  & 0.04  & 0.05  & \  & 0.25  & 0.16 \tabularnewline
\  & S-AMA  & \textbf{0.96}  & 0.06  & \  & 0.03  & 0.07  & \  & 0.30  & 0.21 \tabularnewline
\hline
Setting 2  & k-means  & 0.95  & 0.11  & \  & 0.00  & 0.00  & \  & 1.00  & 0.00 \tabularnewline
\  & ADMM  & 0.14  & 0.20  & \  & 0.00  & 0.00  & \  & 1.00  & 0.00 \tabularnewline
\  & AMA  & 0.08  & 0.21  & \  & 0.00  & 0.00  & \  & 1.00  & 0.00 \tabularnewline
\  & S-AMA  & \textbf{0.97}  & 0.07  & \  & 0.07  & 0.09  & \  & 0.11  & 0.10 \tabularnewline
\hline
Setting 3  & k-means  & 0.83  & 0.15  & \  & 0.00  & 0.00  & \  & 1.00  & 0.00 \tabularnewline
\  & ADMM  & 0.56  & 0.22  & \  & 0.00  & 0.00  & \  & 1.00  & 0.00 \tabularnewline
\  & AMA  & 0.47  & 0.21  & \  & 0.00  & 0.00  & \  & 1.00  & 0.00 \tabularnewline
\  & S-ADMM  & 0.82  & 0.14  & \  & 0.04  & 0.06  & \  & 0.25  & 0.24 \tabularnewline
\  & S-AMA  & \textbf{0.84}  & 0.13  & \  & 0.02  & 0.04  & \  & 0.11  & 0.18 \tabularnewline
\hline
Setting 4  & k-means  & 0.89  & 0.14  & \  & 0.00  & 0.00  & \  & 1.00  & 0.00 \tabularnewline
\  & ADMM  & 0.31  & 0.23  & \  & 0.00  & 0.00  & \  & 1.00  & 0.00 \tabularnewline
\  & AMA  & 0.31  & 0.20  & \  & 0.00  & 0.00  & \  & 1.00  & 0.00 \tabularnewline
\  & S-AMA  & \textbf{0.94}  & 0.09  & \  & 0.01  & 0.02  & \  & 0.01  & 0.03 \tabularnewline
\hline
Setting 5  & k-means  & 0.51  & 0.07  & \  & 0.00  & 0.00  & \  & 1.00  & 0.00 \tabularnewline
\  & ADMM  & 0.54  & 0.08   & \  & 0.00  & 0.00  & \  & 1.00  & 0.00 \tabularnewline
\  & AMA  & 0.53  &  0.09  & \  & 0.00  & 0.00  & \  & 1.00  & 0.00 \tabularnewline
\  & S-AMA  & \textbf{0.57} & 0.07  & \  & 0.00  &  0.00  & \  & 0.34 & 0.27  \tabularnewline
\  & SPECC & 0.52 & 0.08  & \  & 0.00  & 0.00  & \  & 1.00  & 0.00   \tabularnewline
\hline
\end{tabular}
\end{table}

First, convex clustering does not perform well when the feature dimension
is high, even much worse than k-means. Similar phenomenon was also
observed in the simulation studies conducted in \citet{Tan2015}.
This is the motivation for developing sparse convex clustering. Second,
sparse convex clustering improves convex clustering significantly.
Sparse convex clustering (S-AMA) performs as well as k-means when
$p=150$, and performs better than k-means when $p=500$. Third, sparse
convex clustering selects informative feature with great accuracy,
that is, with low FNR and FPR. The feature selection performance of
sparse convex clustering is very promising for settings where $p=500$. Compared with S-ADMM, S-AMA is computationally faster and also delivers slightly better accuracy. Therefore, we recommend
S-AMA in practice. Fourth, for Setting 5, due to its non-spherical essential, convex clustering outperforms k-means. Surprisingly, SPECC does not outstand in terms of the RAND index. The existence of many noninformative features undermines the SPECC and it confirms the necessity of selecting informative features. Sparse convex clustering performs the best by selecting informative features.

\subsection{Application}

\label{sec:real}

We evaluate the performance of sparse convex clustering in LIBRAS
movement data from the Machine Learning Repository \citep{Lichman2013}.
The original dataset contains 15 classes with each class referring
to a hand movement type. Each class contains 24 observations, and
each observation has 90 features consisting of the coordinates of
hand movements. We use this dataset without the clustering assignments
to evaluate each clustering algorithms and then compare the results
with the true classes to compute the RAND index. Before cluster analysis,
each feature is centered. In our S-AMA algorithm, we set $m=5$ and
$\phi=1$ for weight $w_{i_{1},i_{2}}$.

Note that some of the original 15 clusters indicate similar hand movements,
such as curved/vertical swing and horizontal/vertical straight-line.
By plotting the first two principal components of the 90 features,
one can see that some clusters are severely overlapped. Therefore,
for evaluation purpose, six clusters, including vertical swing (labeled as 3),
anti-clockwise arc (labeled as 4), clockwise arc (labeled as 5), horizontal straight-line (labeled as 7),
horizontal wavy (labeled as 11), and
vertical wavy (labeled as 12) in the original dataset are selected. The left panel of Figure \ref{truelabel_cvxpath} displays the
plot of the first principal component (PC1) against the second principal
component (PC2) of 90 features for the selected six clusters with
the true cluster labels.

\begin{figure}[!htb]
\protect\caption{Left panel shows the plot of the first principal component against the second principal component of 90 features for the selected six clusters with true cluster
labels; Right panel shows the clustering path of convex clustering (AMA) using all 90 features by plotting the first principal component (PC1) against the second principal component (PC2).}
\centering \includegraphics[scale=0.8]{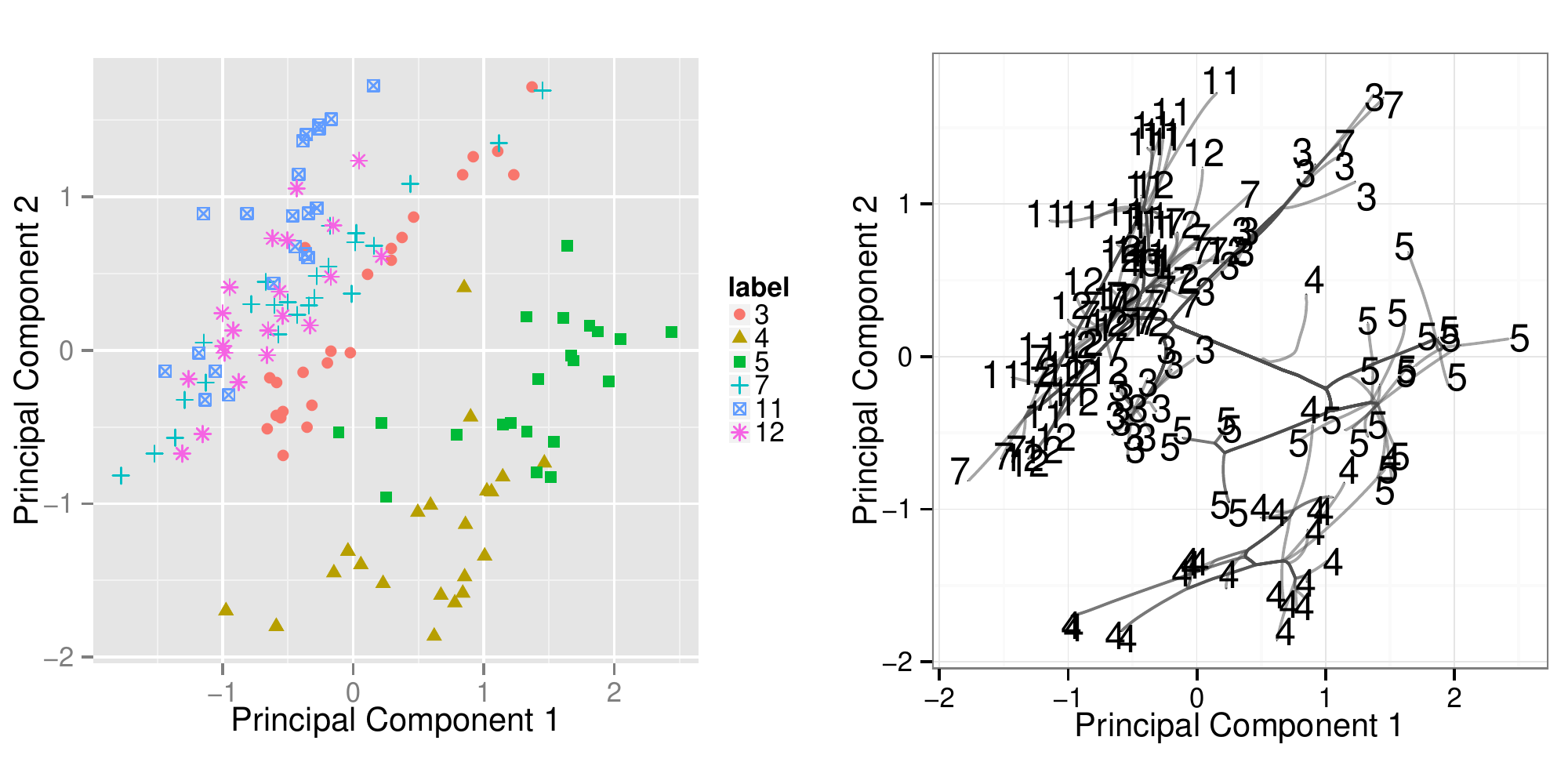} \label{truelabel_cvxpath}
\end{figure}

We first display the clustering path of convex clustering (AMA) using
all 90 features in the right panel of Figure \ref{truelabel_cvxpath}. Clearly, convex clustering
is only able to distinguish clusters 4 and 5 and treat the rest clusters
as one class. This phenomenon shows the curse of dimensionality in
high-dimensional clustering and motivates the need to conduct feature
selection for improved clustering performance.

We use S-AMA to solve sparse convex clustering. The tuning parameters
are selected according to the stability selection in Section \ref{sec:tuning}.
Table \ref{tb_realdata_1} reports the number of estimated clusters,
the number of selected features, and the RAND index between the estimated
cluster membership and the true cluster membership for k-means,
AMA algorithm, and our S-AMA algorithm. Clearly, both convex clustering
(AMA) and sparse convex clustering (S-AMA) perform better than k-means,
which indicates that the performance of convex clustering or sparse
convex clustering is less sensitive to the assumption of spherical
clustering centers. In addition, by using only $13$ informative features,
our S-AMA is able to improve the clustering accuracy of convex clustering
(AMA) by $45\%$. This indicates the importance of variable selection
in high-dimensional clustering.

\begin{table}[!htb]
\centering \protect\caption{The number of estimated clusters, the number of selected features,
and the RAND index for k-means, AMA, and our S-AMA algorithm.}

\label{tb_realdata_1} \vspace{6pt}
 \centering %
\begin{tabular}{c|ccc}
\hline
Algorithm  & \# of clusters  & \# of features  & RAND index \\
\hline
k-means  & 2  & 90  & 0.06 \\
AMA  & 3  & 90  & 0.31 \\
S-AMA  & 3  & 13  & \textbf{0.45}\\
\hline
\end{tabular}
\end{table}

Next we demonstrate the clustering path of sparse convex clustering
(S-AMA) with the 13 selected features in Figure \ref{scvxpath}.
Figure \ref{scvxpath} displays three big clusters, which is consistent
with the number of estimated clusters shown in Table \ref{tb_realdata_1}.
As tuning parameter $\gamma_{1}$ increases,
the clustering path of S-AMA tends to merge clusters 3, 7 and 12 into
one big cluster, merge cluster 4 and 5 into another big cluster, and
identify cluster 11 as the third cluster. This finding is displayed in the final clustering
path of S-AMA executed at the selected $\gamma_{1}$ and $\gamma_{2}$
as shown in Figure \ref{truebestlabel}. In the plot, the left-panel
graph shows the true cluster labels and the right-panel graph shows
the three estimated clusters using S-AMA.

\begin{figure}[!htb]
\protect\caption{The clustering path of sparse convex clustering (S-AMA) using only
13 selected features by plotting PC1 against PC2. These 13 features
are selected via stability selection.}
\centering \includegraphics[scale=0.5]{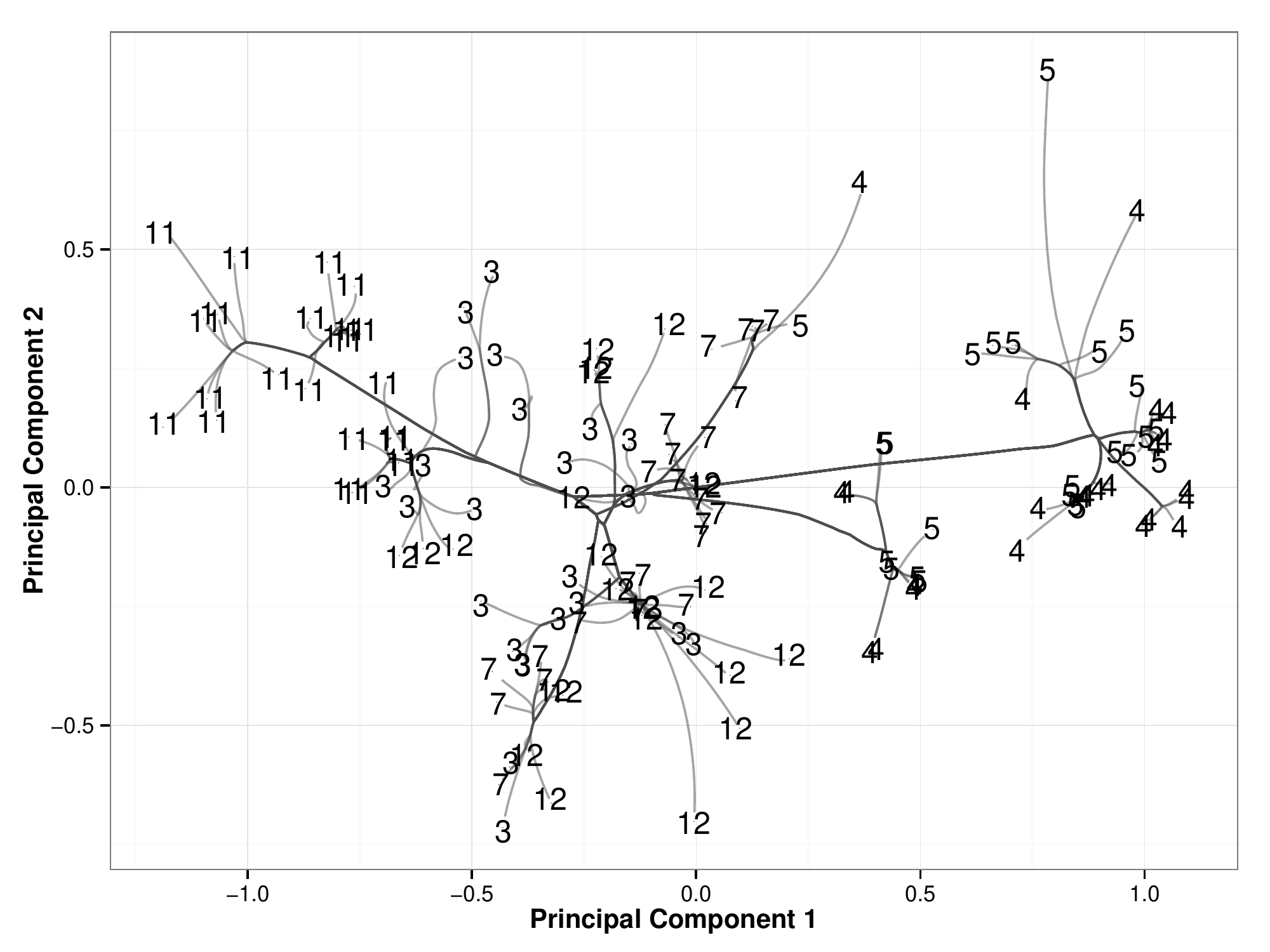}\label{scvxpath}
\end{figure}

\begin{figure}[!htb]
\protect\caption{The left-panel graph shows the true cluster labels by plotting PC1
against PC2 using only 13 selected features. The right-panel graph
shows the estimated cluster membership using S-AMA at the selected
tuning parameters.}
\centering \includegraphics[width=16cm,height=8cm]{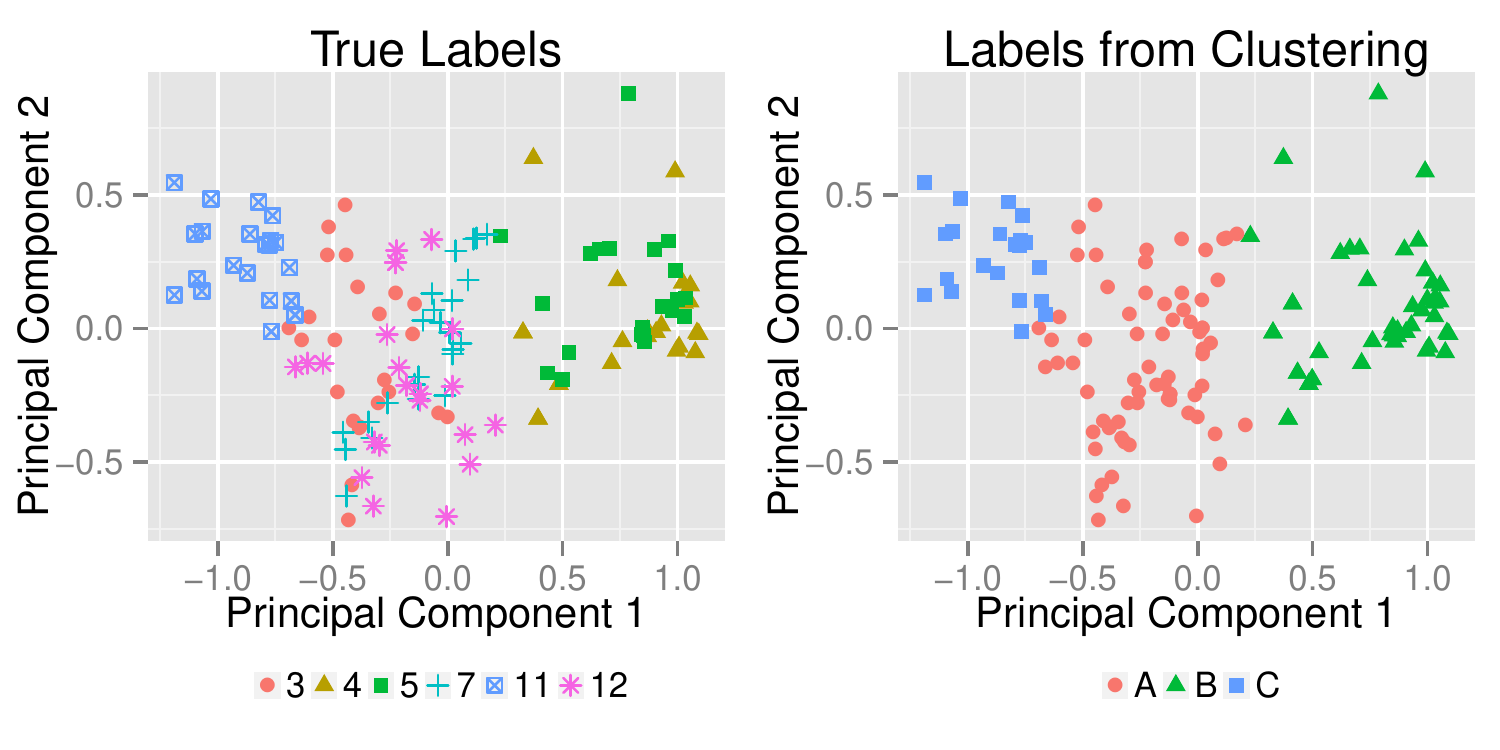}\label{truebestlabel}
\end{figure}

\section{Summary} \label{sec:summary}

In this paper, an extension of convex clustering, sparse convex clustering, is proposed to simultaneously cluster observations and conduct feature selection. Two algorithms, S-ADMM and S-AMA, are developed to implement the new method. The numerical results show that S-AMA is computationally faster and delivers better performance than S-ADMM. In addition, the numerical results show that the selection of tuning parameters in sparse convex clustering is important and the tuning method based on clustering stability performs well. Moreover, this work can motivate future work. \cite{Chi2016convex} presented a convex formulation of the biclustering problem, which seeks to cluster observations and features at the same time. Similarly, we can extend convex biclustering to sparse convex biclustering, in order to conduct biclustering and feature selection simultaneously.

%

\bibliographystyle{apa}
\bibliography{refs_clustering}

\newpage{}

\section*{Appendix}

\setcounter{equation}{0}
\global\long\def\theequation{A.\arabic{equation}}
 \global\long\def\thesubsection{A.\arabic{subsection}}
 \long\def\thelemma{A.\arabic{lemma}}

In Appendix, we provide proofs of Lemma \ref{thm:equivalent} and Proposition \ref{prop1}. All technical details for update steps of S-AMA, proofs of Theorem \ref{thm:bound1}-\ref{thm:selection_consistency}, and degrees of freedom are included in  Supplementary.

\noindent {\bf Proof of Lemma \ref{thm:equivalent}}

Denote $\bfa=\textrm{vec}(\bfA)$, a vectorization of the matrix
$\bfA$. According to the fact that $A_{i_{1}\cdot}-A_{i_{2}\cdot}=\bfA\trans(\bfe_{i_{1}}-\bfe_{i_{2}})$
and the property of the tensor product $\textrm{vec}(\bfR\bfS\bfT)=[\bfT\trans \otimes\bfR]\textrm{vec}(\bfS)$,
solving the minimization of $f(\bfA)$ is equivalent to minimize
\begin{eqnarray*}
f(\bfa)=\frac{1}{2}\|\bfx-\bfa\|_{2}^{2}+\frac{\nu}{2}\sum_{l\in\calE}\|\bfB_{l}\bfP\bfa-\wt{\bfv}_{l}\|_{2}^{2}+\gamma_{2}\sum_{j=1}^{p}u_{j}\|\bfa_{j}\|_{2},
\end{eqnarray*}
where $\bfB_{l}=(\bfe_{i_{1}}-\bfe_{i_{2}})\trans\otimes\bfI_{p}$
and $\bfP$ is a permutation matrix such that $\textrm{vec}(\bfA\trans)=\bfP\textrm{vec}(\bfA)$ ($\bfP\trans=\bfP^{-1}$). Letting
$\bfB\trans=\left(\bfB\trans_{1},\ldots,\bfB\trans_{|\calE|}\right),\wt{\bfv}\trans=\left(\wt{\bfv}_{1}\trans,\ldots,\wt{\bfv}_{|\calE|}\trans\right)$,
it becomes
\begin{eqnarray*}
f(\bfa)=\frac{1}{2}\|\bfx-\bfa\|_{2}^{2}+\frac{\nu}{2}\|\bfB\bfP\bfa-\wt{\bfv}\|_{2}^{2}+\gamma_{2}\sum_{j=1}^{p}u_{j}\|\bfa_{j}\|_{2}.
\end{eqnarray*}
To further simplify the formulae, the following proposition is needed,
with proof shown later.
\begin{proposition} \label{prop1} For a permutation
matrix $\bfP$ such that $\textrm{vec}(\bfA\trans)=\bfP\textrm{vec}(\bfA)$ and any $n$-dim vector $\bfd$, $\left[\bfd\trans\otimes\bfI_{p}\right]\bfP=\bfI_{p}\otimes\bfd\trans.$
\end{proposition}
By Proposition \ref{prop1}, $\bfB_{l}\bfP=\bfI_{p}\otimes(\bfe_{i_{1}}-\bfe_{i_{2}})\trans\triangleq\bfC_{l}$.
Let $\bfC\trans=\left(\bfC\trans_{1},\ldots,\bfC\trans_{|\calE|}\right)$,
then the second term in $f(\bfa)$ becomes $\frac{\nu}{2}\|\bfC\bfa-\wt{\bfv}\|_{2}^{2}=\frac{\nu}{2}\sum_{j=1}^{p} \sum_{l\in\calE}\left((\bfe_{i_{1}}-\bfe_{i_{2}})\trans\bfa_{j}-\wt{v}_{jl}\right)_{2}^{2}$.

Therefore, the objective function can be separated to $p$ sub-optimization
questions:
\begin{eqnarray*}
\min_{\bfa_{j}}\frac{1}{2}\|\bfx_{j}-\bfa_{j}\|_{2}^{2}+\frac{\nu}{2}\sum_{l\in\calE}\left((\bfe_{i_{1}}-\bfe_{i_2})\trans\bfa_{j}-\wt{v}_{jl}\right)_{2}^{2} +\gamma_{2}u_{j}\|\bfa_{j}\|_{2},\quad j=1,\ldots,p.
\end{eqnarray*}
By some algebra, if $\calE$ contains all possible edges, it can be
rewritten as
\begin{eqnarray}
\min_{\bfa_{j}}\frac{1}{2}\bfa_{j}\trans\bfM\bfa_{j}-\bfz_{i}\trans\bfa_{j} +\frac{1}{2}\nu\wt{\bfv}_{j\cdot}\trans\wt{\bfv}_{j\cdot}+\gamma_{2}u_{j}\|\bfa_{j}\|_{2},\label{subobj}
\end{eqnarray}
where $\wt{\bfv}_{j\cdot}=(\wt{v}_{j1},\ldots,\wt{v}_{j|\calE|})\trans$,
$\bfM=\bfI_{n}+\nu\sum_{l\in\calE}(\bfe_{i_{1}}-\bfe_{i_{2}})(\bfe_{i_{1}}-\bfe_{i_{2}})\trans=(1+n\nu)\bfI_{n}-\nu\bfone_{n}\bfone_{n}\trans$,
and $\bfz_{j}=\bfx_{j}+\nu\sum_{l\in\calE}\wt{v}_{jl}(\bfe_{i_{1}}-\bfe_{i_{2}})$.
The KKT conditions of \eqref{subobj} are
\begin{eqnarray*}
\forall\bfa_{j}\neq\bfzero,\quad\bfM\bfa_{j}-\bfz_{i}+\frac{\gamma_{2}u_{j}\bfa_{j}}{\|\bfa_{j}\|_{2}}=\bfzero;\quad\quad\ \forall\bfa_{j}=\bfzero,\quad\|\bfz_{j}\|_{2}\leq\gamma_{2}u_{j}.
\end{eqnarray*}

Here are some remarks on the above KKT conditions. $\bfM$ is positive
definite, and thus it can be diagonalized by $\bfM=\bfS\trans\bPhi\bfS$,
where $\bPhi=\textrm{diag}(\phi_{1},\ldots,\phi_{n})$ and $\bfS$
is an orthogonal matrix. It can be verified that $\phi_{1}=1$ and
$\phi_{i}=1+n\nu,i=2,\ldots,n$. Then $\bPhi\bfS\bfa_{j}-\bfS\bfz_{j}+\frac{\gamma_{2}u_{j}\bfS\bfa_{j}}{\|\bfS\bfa_{j}\|_{2}}=\bfzero$.
Let $\wt\bfa_{j}=\bfS\bfa_{j}$. One needs to solve $\bPhi\wt\bfa_{j}-\bfS\bfz_{j}+\frac{\gamma_{2}u_{j}\wt\bfa_{j}}{\|\wt\bfa_{j}\|_{2}}=\bfzero$.
If $\nu=0$, then $\bPhi=\bfI$, implying the solution $\wt\bfa_{j}$ shares the same
direction with $\bfz_j$. In this case, an explicit soft-threshold formula can be obtained,
and this situation under the standard group LASSO problem was discussed
in \citet{Yuan2006}. But if $\nu\neq0$, a scaling transformation
is applied to $\wt\bfa_{j}$, and there is no explicit solution.

Alternatively, we can rewrite \eqref{subobj} so that existing algorithms
can be applied. Define $\bfN=\sqrt{1+n\nu}\bfI_{n}-\frac{\sqrt{1+n\nu}-1}{n}\bfone_{n}\bfone_{n}\trans$,
which performs like a ``design matrix". It can be verified
that $\bfM=\bfN\bfN$ and $\bfN^{-1}$ has the form defined in Lemma \ref{thm:equivalent}. Let $\bfy_{j}=(\bfN)^{-1}\bfz_{j}$, which
performs like a pseudo outcome in the $j$-th sub-problem. Then \eqref{subobj}
is equivalent to $\min_{\bfa_{j}}\frac{1}{2}\|\bfy_{j}-\bfN\bfa_{j}\|_{2}^{2}+\gamma_{2}u_{j}\|\bfa_{j}\|_{2}$.
Note that during the whole algorithm, $\bfN$ and its inverse are
calculated only once. This ends the proof of Lemma \ref{thm:equivalent}.
\hfill{}$\blacksquare$

\noindent \textbf{{Proof of Proposition \ref{prop1}:}} Note that
$\bfP=(P_{kl}),1\leq k,l\leq np$ here is a unique permutation matrix
such that $P_{kl}=1$ if $k=(i-1)p+j$ and $l=(j-1)n+i,1\leq i\leq n,1\leq j\leq p$,
and 0 otherwise. By the definition of $\bfP$, it is clear that multiplying
a matrix by $\bfP$ on the right moves its $k$-th column to the $l$-th
column when $P_{kl}=1$.

Consider the $i$-th element $d_{i}$ of $\bfd$, then in $\bfd\trans\otimes\bfI_{p}$,
its entries at $(j,(i-1)p+j)$ equal $d_{i}$, $j=1,\ldots,p$. Thus,
in $(\bfd\trans\otimes\bfI_{n})\bfP$, the entry at $(j,(j-1)n+i)$
equals to $d_{i}$. In $\bfI_{p}\otimes\bfd\trans$, it is easy to
see the entry at $(j,(j-1)n+i)$ equal $d_{i},i=1,\ldots,n,j=1,\ldots,p$. \hfill{}$\blacksquare$

\newpage

\pagenumbering{arabic}
\setcounter{page}{1}
\baselineskip=14pt

\begin{center}
{\Large \tits.arg} \\

\vskip 3mm

\author.arg
\end{center}

\footnotetext
{ Correspondence to: 650 First Avenue Rm 578, New York, NY 10016; Email: Binhuan.Wang@nyumc.org}

\vskip 3mm

\date{}

\setcounter{equation}{0}
\setcounter{lemma}{0}
\global\long\def\theequation{S.\arabic{equation}}
 \global\long\def\thesubsection{S.\arabic{subsection}}
 \long\def\thelemma{S.\arabic{lemma}}

\noindent In this supplementary, we provide all technical details for update steps of S-AMA, proofs of Theorems \ref{thm:bound1}-\ref{thm:selection_consistency} and degrees of freedom.

\subsection*{S.1 Update Steps of S-AMA}

By letting $\nu=0$ while updating $\bfA$, the S-ADMM algorithm can
be simplified significantly. Noting that $\bfM=\bfI_{n}$ and $\bfz_{j}=\bfx_{j}+\sum_{l\in\calE}\lambda_{jl}(\bfe_{i_{1}}-\bfe_{i_{2}})$,
where $\lambda_{jl}$ is the $j$-th element of $\blambda_{l}$, the
KKT conditions are
\[
\forall\bfa_{j}\neq\bfzero,\quad\bfa_{j}-\bfz_{j}+\frac{\gamma_{2}u_{j}\bfa_{j}}{\|\bfa_{j}\|}=\bfzero;\quad\quad\forall\bfa_{j}=\bfzero,\quad\|\bfz_{i}\|_{2}\leq\gamma_{2}u_{j}.
\]
The solutions are $\wh\bfa_{j}=\left(1-\frac{\gamma_{2}u_{j}}{\|\bfz_{j}\|_{2}}\right)_{+}\bfz_{j}$,
where $(z)_{+}=\max\{0,z\}$. See \citet{Yuan2006}.

By applying the projection method, one can update $\bfv_{l}$ and
$\blambda_{l}$ as $\bfv_{l}^{m+1}=A_{i_{1}\cdot}^{m+1}-A_{i_{2}\cdot}^{m+1}-\nu^{-1}\blambda_{l}^{m}-\calP_{tB}[A_{i_{1}\cdot}^{m+1}-A_{i_{2}\cdot}^{m+1}-\nu^{-1}\blambda_{l}^{m}]$,
where $t=\sigma_{l}=\gamma_{1}w_{l}/\nu$ and $\calP_{B}(\bfz)$ denotes
projection onto $B=\{\bfy:\|\bfy\|_{\dag}\leq1\}$. The point $\calP_{B}(\bfz)$
can be characterized by the relations $\calP_{B}(\bfz)\in B$ and
$\forall\bfy\in B,\langle\bfy-\calP_{B}(\bfz),\bfz-\calP_{B}(\bfz)\rangle\leq0$.
Then
\begin{eqnarray*}
\blambda_{l}^{m+1} & = & \blambda_{l}^{m}+\nu(\bfv_{l}^{m+1}-A_{i_{1}\cdot}^{m+1}+A_{i_{2}\cdot}^{m+1})\\
 & = & -\nu\calP_{tB}(A_{i_{1}\cdot}^{m+1}-A_{i_{2}\cdot}^{m+1}-\nu^{-1}\blambda_{1l}^{m}) \\
 & = & \calP_{C_{l}}(\blambda_{l}^{m}-\nu\bfg_{l}^{m+1}),
\end{eqnarray*}
where $\bfg_{l}^{m}=A_{i_{1}\cdot}^{m}-A_{i_{2}\cdot}^{m}$ and $C_{l}=\{\blambda_{l}:\|\blambda_{l}\|_{\dag}\leq\gamma_{1}w_{l}\}$.
Note that there is no need to update $\bfv_{l}$, and $\blambda_{l}$
can be directly updated.

\subsection*{S.2 Proofs of Theorems \ref{thm:bound1}-\ref{thm:selection_consistency}}

Before we prove the main theorems, we introduce the following lemma to bound the quadratic forms of independent sub-Gaussian random variables. This is a standard result in \cite{Hanson1971}.
\begin{lemma} \label{lemma:HW71}
Let $\bepsilon$ be a $d$-dim vector of independent sub-Gaussian random variables with mean zero and variance $\sigma^2$. Let $\bf A$ be a symmetric matrix. For any $t>0$, there exists some positive constants $c_1,c_2$ such that,
\begin{eqnarray*}
P\left( \bepsilon^{\top}\bf A \bepsilon  > t  + \sigma^2 \textrm{tr}(\bf A)\right) \leq \exp \left\{ - \min\left(  \frac{c_1t^2}{\sigma^2 \| {\bf A} \|_F}, \frac{c_2t}{\sigma^2 \| {\bf A} \|_{sp}} \right)\right\}
\end{eqnarray*}
\end{lemma}

In addition, we need the following lemma for tail probability of a linear combination of independent sub-Gaussian random variables. This is a standard result in \cite{Rigollet2015}.

\begin{lemma} \label{lemma:subGausumtail}
Let $\bepsilon$ be a $d$-dim vector of independent sub-Gaussian random variables with mean zero and variance $\sigma^2$. Then for any $\bfb \in \real^d$, we have
\begin{eqnarray*}
P\left( \bfb\trans\bepsilon > t \right) \leq \exp \left( -\frac{t^2}{2\sigma^2 \| \bfb \|_2^2}\right) \quad
{\rm and} \quad
P\left( \bfb\trans\bepsilon < -t \right) \leq \exp \left(- \frac{t^2}{2\sigma^2 \| \bfb \|_2^2}\right).
\end{eqnarray*}
\end{lemma}

We apply the matrix decomposition proposed by \cite{Liu2013} to simplify the analysis. According to Lemma 1 in \cite{Tan2015}, the rank of $\bfC$ is $p(n-1)$. Let $\bfC=\bfR\bfD \bfS_{\psi}\trans$ be the singular value decomposition of $\bfC$, where $\bfR \in \real^{\left[p \binom{n}{2}\right] \times p(n-1)}$ such that $\bfR\trans \bfR=\bfI$, $\bfD \in \real^{p(n-1) \times p(n-1)}$ is a diagonal matrix, and $\bfS_{\psi} \in \real^{np \times p(n-1)}$ such that $\bfS_{\psi}\trans \bfS_{\psi} =\bfI$. There must exist a matrix $\bfS_{\phi} \in \real^{np \times p}$ such that $\bfS=[\bfS_{\phi},\bfS_{\psi}] \in \real^{np \times np}$ is an orthogonal matrix. Then it is clear that $\bfS_{\phi}\trans \bfS_{\psi}=\bfzero$.

Additionally, let $\bphi=\bfS_{\phi}\trans\bfa \in \real^{p}$ and $\bpsi=\bfS_{\psi}\trans \bfa \in \real^{p(n-1)}$, and thus $\bfa=\bfS_{\phi}\bphi + \bfS_{\psi}\bpsi$. Similarly, we define $\bphi_0=\bfS_{\phi}\trans\bfa_0$ and $\bpsi_0=\bfS_{\psi}\trans \bfa_0$, and thus $\bfa_0=\bfS_{\phi}\bphi_0 + \bfS_{\psi}\bpsi_0$.

Therefore, we can rewrite the above minimization problem as follows:
\begin{eqnarray} \label{eq:phipsi}
\min_{\bphi, \bpsi} \frac{1}{2} \| \bfx - \bfS_{\phi}\bphi -\bfS_{\psi} \bpsi \|_2^2 + \gamma_1 \sum_{l \in \calE} \|\bfG_l \bpsi \|_q + \gamma_2 \sum_{j=1}^{p} u_j \| \bfE_j (\bfS_{\phi}\bphi + \bfS_{\psi}\bpsi) \|_2.
\end{eqnarray}
where $\bfE_j=\bfe_j^{*T} \otimes \bfI_n$ and $\bfG_l$ is a submatrix of $\bfG=(\bfG_1\trans,\ldots,\bfG_{|\calE |}\trans)\trans$ such that $\bfG=\bfR \bfD$.
Note that the rank of $\bfG$ is $p(n-1)$. Thus there exists the Moore-Penrose pseudo-inverse $\bfG^{+} \in \real^{p(n-1) \times \left[ p \cdot \binom n2\right]}$ such that $\bfG^+ \bfG=\bfI$. Let $\wh\bphi$ and $\wh\bpsi$ are the solution to \eqref{eq:phipsi}. Then it is natural to see  $\wh\bphi=\bfS_{\phi}\trans\wh\bfa$, $\wh\bpsi=\bfS_{\psi}\trans \wh\bfa$ and thus $\wh\bfa=\bfS_{\phi}\wh\bphi + \bfS_{\psi}\wh\bpsi$.

\noindent\textbf{Proof of Theorem \ref{thm:bound1}:} $\wh\bphi$ and $\wh\bpsi$ are the minimizers to \eqref{eq:phipsi}, so we know
\begin{eqnarray*}
&&\frac{1}{2} \| \bfx - \bfS_{\phi}\wh\bphi -\bfS_{\psi} \wh\bpsi \|_2^2 + \gamma_1  \|\bfG \wh\bpsi \|_1 + \gamma_2 \sum_{j=1}^{p} u_j \| \bfE_j (\bfS_{\phi}\wh\bphi + \bfS_{\psi}\wh\bpsi) \|_2 \\
 && \leq \frac{1}{2} \| \bfx - \bfS_{\phi}\bphi_0 -\bfS_{\psi} \bpsi_0 \|_2^2 + \gamma_1  \|\bfG \bpsi_0 \|_1 + \gamma_2 \sum_{j=1}^{p} u_j \| \bfE_j (\bfS_{\phi}\bphi_0 + \bfS_{\psi}\bpsi_0) \|_2.
\end{eqnarray*}
By some algebra, it follows
\begin{eqnarray} \label{eq:q1}
 \frac{1}{2} \left\| \bfS_{\phi}(\wh\bphi - \bphi_0) + \bfS_{\psi} (\wh\bpsi - \bpsi_0) \right\|^2_2
\leq g(\wh\bphi, \wh\bpsi) + \gamma_1 \left(\|\bfG \bpsi_0 \|_1 - \|\bfG \wh\bpsi \|_1 \right) \nonumber \\
+ \gamma_2 \sum_{j=1}^p u_j \left[
\| \bfE_j (\bfS_{\phi}\bphi_0 + \bfS_{\psi}\bpsi_0) \|_2 - \| \bfE_j (\bfS_{\phi}\wh\bphi + \bfS_{\psi}\wh\bpsi) \|_2 \right],
\end{eqnarray}
where $g(\wh\bphi, \wh\bpsi)=\bvarepsilon\trans\big[ \bfS_{\phi}(\wh\bphi - \bphi_0) + \bfS_{\psi} (\wh\bpsi - \bpsi_0) \big]$.
Next, we build up the relation between $\wh\bphi$ and $\bphi_0$. The optimal condition for $\wh\bphi$ based on \eqref{eq:phipsi} is
\begin{eqnarray*}
-\bfS_{\phi}\trans (\bfx - \bfS_{\phi}\wh\bphi - \bfS_{\psi} \wh\bpsi) + \bxi = \bfzero,
\end{eqnarray*}
where $\bxi=\gamma_2 \sum_{j=1}^p u_j \frac{\bfS_{\phi}\trans \bfE_j\trans \bfE_j(\bfS_{\phi} \wh\bphi + \bfS_{\psi}\wh\bpsi)} { \| \bfE_j(\bfS_{\phi} \wh\bphi + \bfS_{\psi}\wh\bpsi) \|_2}$. Then we have $\wh\bphi - \bphi_0 = \bfS_{\phi}\trans \bvarepsilon - \bxi$. Thus,
\begin{eqnarray*}
\frac{1}{np} \Big| g(\wh\bphi,\wh\bpsi) \Big|
&=& \frac{1}{np} \Big| \bvarepsilon\trans \big[ \bfS_{\phi}(\bfS_{\phi}\trans \bvarepsilon - \bxi) + \bfS_{\psi}(\wh\bpsi - \bpsi_0) \big] \Big| \\
& \leq & \frac{1}{np} \Big| \bvarepsilon\trans \bfS_{\phi}\bfS_{\phi}\trans \bvarepsilon \Big| +\frac{1}{np} \Big|  \bvarepsilon\trans \bfS_{\phi}\bxi \Big|+ \frac{1}{np} \Big|\bvarepsilon\trans\bfS_{\psi}(\wh\bpsi - \bpsi_0) \Big| \\
& \leq & \frac{1}{np} \Big| \bvarepsilon\trans \bfS_{\phi}\bfS_{\phi}\trans \bvarepsilon \Big| +\frac{1}{np} \Big|  \bvarepsilon\trans \bfS_{\phi}\bxi \Big|+ \frac{1}{np} \| \bvarepsilon\trans\bfS_{\psi} \bfG^{+} \|_{\infty} \| \bfG(\wh\bpsi - \bpsi_0)\|_1.
\end{eqnarray*}
The last inequality follows from the fact $\bfG^{+} \bfG = \bfI$ and the H\"{o}lder's inequality. Now we need to establish bounds for the above three items on the right-hand side of the inequality.

\textbf{Bounds for $\frac{1}{np} \bvarepsilon\trans \bfS_{\phi}\bfS_{\phi}\trans \bvarepsilon$ and $\| \bvarepsilon\trans\bfS_{\psi} \bfG^{+} \|_{\infty} $}:

Based on arguments used for Lemma 6 in \cite{Tan2015}, if follows
\begin{eqnarray} \label{eq:bound1}
P\left( \frac{1}{np} \bvarepsilon\trans \bfS_{\phi}\bfS_{\phi}\trans \bvarepsilon  \geq \sigma^2 \left[ \frac{1}{n}+\sqrt{\frac{\log (np)}{n^2p}}\right] \right)
\leq \exp \big\{ - \min \big( c_1 \log (np), c_2 \sqrt{p \log (np)} \big) \big\},
\end{eqnarray}
and
\begin{eqnarray} \label{eq:bound2}
P \left( \| \bvarepsilon\trans\bfS_{\psi} \bfG^{+} \|_{\infty} \geq 2\sigma \sqrt{\frac{\log (p \cdot \binom n2 )}{n}} \right)
\leq \frac{2}{p\cdot \binom n2}.
\end{eqnarray}

\textbf{Bound for $\frac{1}{np} \Big|  \bvarepsilon\trans \bfS_{\phi}\bxi \Big|$}:

Note that $\| \bfS_{\phi} \bxi \|_2 \leq \| \bfS_{\phi} \|_2 \| \bxi \|_2 = \| \bxi \|_2$ because $\| \bfS_{\phi} \|_2=1$. Next, we have
\begin{eqnarray*}
\| \bxi \|_2 &=& \left\| \gamma_2 \sum_{j=1}^p u_j \frac{\bfS_{\phi}\trans \bfE_j\trans \bfE_j(\bfS_{\phi} \wh\bphi + \bfS_{\psi}\wh\bpsi)} { \| \bfE_j(\bfS_{\phi} \wh\bphi + \bfS_{\psi}\wh\bpsi) \|_2} \right\|_2 \\
&\leq & \gamma_2 \sum_{j=1}^p u_j \left\|  \bfS_{\phi}\trans \bfE_j\trans \right\|_2 \\
&\leq & \gamma_2 \sum_{j=1}^p u_j \big\| \bfE_j\trans \big\|_2.
\end{eqnarray*}
Note the fact $\bfe_j^{*T} \otimes \bfI_n$, and then $\big\| \bfE_j\trans \big\|_2=1$. Thus, $\| \bfS_{\phi}\bxi \|_2^2 \leq \gamma_2^2 \| \bfu \|_1^2 $. Based on Lemma \ref{lemma:subGausumtail}, we have
\begin{eqnarray*}
P \bigg( \frac{1}{np} \Big|  \bvarepsilon\trans \bfS_{\phi}\bxi \Big| > t \bigg)
\leq 2 \exp \left( -\frac{n^2p^2 t^2}{2 \sigma^2 \| \bfS_{\phi} \bxi \|_2^2} \right)
\leq 2 \exp \left( -\frac{n^2p^2 t^2}{2 \sigma^2 \gamma_2^2  \| \bfu \|_1^2} \right).
\end{eqnarray*}
Setting $t=1/(np)$, it follows
\begin{eqnarray} \label{eq:bound3}
P \bigg( \frac{1}{np} \Big|  \bvarepsilon\trans \bfS_{\phi}\bxi \Big| > \frac{1}{np} \bigg)
\leq 2 \exp \left( -\frac{np}{2 \sigma^2 \gamma_2^2 \| \bfu \|_1^2} \right).
\end{eqnarray}

Therefore, by combining \eqref{eq:bound1}-\eqref{eq:bound3} and setting $\gamma_1 > 4 \sigma \sqrt{\frac{\log (p \cdot \binom{n}{2})}{n}}$, we have that
\begin{eqnarray} \label{eq:bound_g1}
\frac{1}{np}  g(\wh\bphi,\wh\bpsi) \leq \frac{\gamma_1}{2np}\| \bfG(\wh\bpsi - \bpsi_0)\|_1 + \sigma^2 \left[
\frac{1}{n} + \sqrt{\frac{\log (np)}{n^2 p}} \right] + \frac{1}{np}
\end{eqnarray}
holds with probability at least $1-c_3$.

Furthermore, it is clear that
\begin{eqnarray*}
\gamma_2 \sum_{j=1}^p u_j \left[
\| \bfE_j (\bfS_{\phi}\bphi_0 + \bfS_{\psi}\bpsi_0) \|_2 - \| \bfE_j (\bfS_{\phi}\wh\bphi + \bfS_{\psi}\wh\bpsi) \|_2 \right]
&=& \gamma_2\sum_{j=1}^p u_j \left(\| \bfa_{0j} \|_2 - \| \wh\bfa_j \|_2 \right) \\
& \leq & \gamma_2 \sum_{j=1}^p u_j \| \bfa_{0j}  -  \wh\bfa_j \|_2 \\
& \leq & \gamma_2 \| \bfu \|_2 \| \bfa_{0}  -  \wh\bfa \|_2 \\
& \leq & \frac{\gamma_2}{2} \left(\| \bfu \|_2^2 + \| \bfa_{0}  -  \wh\bfa \|_2^2 \right),
\end{eqnarray*}
by noting that $\sum_{j=1}^p \| \bfa_{0j} - \wh\bfa_j \|_2^2 = \| \bfa_0 - \wh\bfa \|_2^2$. Substituting the above inequality and \eqref{eq:bound_g1} into \eqref{eq:q1}, we obtain that
\begin{eqnarray*}
&& \frac{1}{2np} \left\| \bfS_{\phi}(\wh\bphi - \bphi_0) + \bfS_{\psi} (\wh\bpsi - \bpsi_0) \right\|^2_2  \\
&\leq& \frac{\gamma_1}{2np}\| \bfG(\wh\bpsi - \bpsi_0)\|_1 + \sigma^2 \left[
\frac{1}{n} + \sqrt{\frac{\log (np)}{n^2 p}} \right] + \frac{1}{n^2 p^2} \\
 && + \frac{\gamma_1}{np} \left(\|\bfG \bpsi_0 \|_1 - \|\bfG \wh\bpsi \|_1 \right)
  + \frac{\gamma_2}{2np} \left(\| \bfu \|_2^2 + \| \bfa_{0}  -  \wh\bfa \|_2^2 \right)
\end{eqnarray*}
Therefore, it implies that
\begin{eqnarray*}
\frac{1-\gamma_2}{2np}  \| \wh\bfa - \bfa_0 \|_2^2  \leq
\frac{3\gamma_1}{2np} \| \bfC \bfa_0 \|_1 + \frac{\gamma_2 \| \bfu \|_2^2 }{2np} +\sigma^2 \left[
\frac{1}{n} + \sqrt{\frac{\log (np)}{n^2 p}} \right]  + \frac{1}{n^2 p^2}
\end{eqnarray*}
holds with probability at least $1-c_3$. \hfill $\blacksquare$

\noindent\textbf{Proof of Theorem \ref{thm:bound2}:} By the definition of $\wh\bphi$ and $\wh\bpsi$, we know
\begin{eqnarray*}
&&\frac{1}{2} \| \bfx - \bfS_{\phi}\wh\bphi -\bfS_{\psi} \wh\bpsi \|_2^2 + \gamma_1 \sum_{l \in \calE} \|\bfG_l \wh\bpsi \|_2 + \gamma_2 \sum_{j=1}^{p} u_j \| \bfE_j (\bfS_{\phi}\wh\bphi + \bfS_{\psi}\wh\bpsi) \|_2 \\
 && \leq \frac{1}{2} \| \bfx - \bfS_{\phi}\bphi_0 -\bfS_{\psi} \bpsi_0 \|_2^2 + \gamma_1 \sum_{l \in \calE} \|\bfG_l \bpsi_0 \|_2 + \gamma_2 \sum_{j=1}^{p} u_j \| \bfE_j (\bfS_{\phi}\bphi_0 + \bfS_{\psi}\bpsi_0) \|_2,
\end{eqnarray*}
implying
\begin{eqnarray} \label{eq:q2}
&& \frac{1}{2} \left\| \bfS_{\phi}(\wh\bphi - \bphi_0) + \bfS_{\psi} (\wh\bpsi - \bpsi_0) \right\|^2_2 \nonumber \\
&\leq& g(\wh\bphi, \wh\bpsi) + \gamma_1 \sum_{l \in \calE} \left(\|\bfG_l \bpsi_0 \|_2 - \|\bfG_l \wh\bpsi \|_2 \right) \nonumber \\
&& + \gamma_2 \sum_{j=1}^p u_j \left[
\| \bfE_j (\bfS_{\phi}\bphi_0 + \bfS_{\psi}\bpsi_0) \|_2 - \| \bfE_j (\bfS_{\phi}\wh\bphi + \bfS_{\psi}\wh\bpsi) \|_2 \right].
\end{eqnarray}
Next, we build up the relation between $\wh\bphi$ and $\bphi_0$. Following the same arguments used in the case with $q=1$, we have $\wh\bphi - \bphi_0 = \bfS_{\phi}\trans \bvarepsilon - \bxi$. Thus,
\begin{eqnarray*}
\frac{1}{np} \Big| g(\wh\bphi,\wh\bpsi) \Big|
&=& \frac{1}{np} \Big| \bvarepsilon\trans \bfS_{\phi}\bfS_{\phi}\trans \bvarepsilon - \bvarepsilon\trans \bfS_{\phi}\bxi + \bvarepsilon\trans\bfS_{\psi}(\wh\bpsi - \bpsi_0) \Big| \\
& \leq & \frac{1}{np} \Big| \bvarepsilon\trans \bfS_{\phi}\bfS_{\phi}\trans \bvarepsilon \Big| +\frac{1}{np} \Big|  \bvarepsilon\trans \bfS_{\phi}\bxi \Big|+ \frac{1}{np} \Big|\bvarepsilon\trans\bfS_{\psi}(\wh\bpsi - \bpsi_0) \Big| \\
& = & \frac{1}{np} \Big| \bvarepsilon\trans \bfS_{\phi}\bfS_{\phi}\trans \bvarepsilon \Big| +\frac{1}{np} \Big|  \bvarepsilon\trans \bfS_{\phi}\bxi \Big|+ \frac{1}{np} \sum_{l \in \calE}\Big|\bvarepsilon\trans\bfS_{\psi} \bfG^{+}_l \bfG_l(\wh\bpsi - \bpsi_0) \Big| \\
& \leq & \frac{1}{np} \Big| \bvarepsilon\trans \bfS_{\phi}\bfS_{\phi}\trans \bvarepsilon \Big| +\frac{1}{np} \Big|  \bvarepsilon\trans \bfS_{\phi}\bxi \Big|+ \frac{1}{np} \sum_{l \in \calE}\big\|\bvarepsilon\trans\bfS_{\psi} \bfG^{+}_l \big\|_2 \big\| \bfG_l(\wh\bpsi - \bpsi_0) \big\|_2 \\
&\leq &  \frac{1}{np} \Big| \bvarepsilon\trans \bfS_{\phi}\bfS_{\phi}\trans \bvarepsilon \Big| +\frac{1}{np} \Big|  \bvarepsilon\trans \bfS_{\phi}\bxi \Big|+ \frac{1}{np} \max_{l \in \calE} \big\|\bvarepsilon\trans\bfS_{\psi} \bfG^{+}_l \big\|_2\sum_{l \in \calE} \big\| \bfG_l(\wh\bpsi - \bpsi_0) \big\|_2.
\end{eqnarray*}
Now we need to establish bounds for the above three items on the right-hand side of the inequality.

\textbf{Bounds for $\frac{1}{np} \bvarepsilon\trans \bfS_{\phi}\bfS_{\phi}\trans \bvarepsilon$ and $\max_{l \in \calE} \big\|\bvarepsilon\trans\bfS_{\psi} \bfG^{+}_l \big\|_2$}:

Based on the previous arguments used for $q = 1$, if follows
\begin{eqnarray} \label{eq:bound4}
P\bigg( \frac{1}{np} \bvarepsilon\trans \bfS_{\phi}\bfS_{\phi}\trans \bvarepsilon  \geq \sigma^2 \left[ \frac{1}{n}+\sqrt{\frac{\log (np)}{n^2p}}\right] \bigg)
\leq \exp \big\{ - \min \big( c_1 \log (np), c_2 \sqrt{p \log (np)} \big) \big\},
\end{eqnarray}
and Lemma 7 in \cite{Tan2015} shows
\begin{eqnarray} \label{eq:bound5}
P \left( \max_{l \in \calE} \big\|\bvarepsilon\trans\bfS_{\psi} \bfG^{+}_l \big\|_2 \geq 2\sigma \sqrt{\frac{\log (p \cdot \binom n2 )}{n}} \right)
\leq \frac{2}{p\cdot \binom n2}.
\end{eqnarray}

\textbf{Bound for $\frac{1}{np} \Big|  \bvarepsilon\trans \bfS_{\phi}\bxi \Big|$}:

We have shown in the case with $q=1$ that
\begin{eqnarray} \label{eq:bound6}
P \bigg( \frac{1}{np} \Big|  \bvarepsilon\trans \bfS_{\phi}\bxi \Big| > \frac{1}{n p} \bigg)
\leq 2 \exp \left( -\frac{np}{2 \sigma^2 \gamma_2^2 \| \bfu \|_1^2} \right).
\end{eqnarray}

Therefore, by combining \eqref{eq:bound4}-\eqref{eq:bound6} and setting $\gamma_1 > 4 \sigma \sqrt{\frac{\log (p \cdot \binom{n}{2})}{n}}$, we have that
\begin{eqnarray} \label{eq:bound_g2}
\frac{1}{np}  g(\wh\bphi,\wh\bpsi) \leq \frac{\gamma_1}{2np} \sum_{l \in \calE} \| \bfG_l(\wh\bpsi - \bpsi_0)\|_2 + \sigma^2 \left[
\frac{1}{n} + \sqrt{\frac{\log (np)}{n^2 p}} \right] + \frac{1}{n p}
\end{eqnarray}
holds with probability at least $1-c_3$.

Furthermore, we have the following results from the case with $q=1$:
\begin{eqnarray*}
\gamma_2 \sum_{j=1}^p u_j \left[
\| \bfE_j (\bfS_{\phi}\bphi_0 + \bfS_{\psi}\bpsi_0) \|_2 - \| \bfE_j (\bfS_{\phi}\wh\bphi + \bfS_{\psi}\wh\bpsi) \|_2 \right]
 \leq  \frac{\gamma_2}{2} \left(\| \bfu \|_2^2 + \| \bfa_{0}  -  \wh\bfa \|_2^2 \right).
\end{eqnarray*}

Substituting the above inequality and \eqref{eq:bound_g2} into \eqref{eq:q2}, we obtain that
\begin{eqnarray*}
&& \frac{1}{2np} \left\| \bfS_{\phi}(\wh\bphi - \bphi_0) + \bfS_{\psi} (\wh\bpsi - \bpsi_0) \right\|^2_2  \\
&\leq& \frac{\gamma_1}{2np} \sum_{l\in\calE}\| \bfG_l(\wh\bpsi - \bpsi_0)\|_2 + \sigma^2 \left[
\frac{1}{n} + \sqrt{\frac{\log (np)}{n^2 p}} \right] + \frac{1}{n^2 p^2} \\
 && + \frac{\gamma_1}{np} \sum_{l\in\calE} \left(\|\bfG_l \bpsi_0 \|_2 - \|\bfG_l \wh\bpsi \|_2 \right)
  + \frac{\gamma_2}{2np} \left(\| \bfu \|_2^2 + \| \bfa_{0}  -  \wh\bfa \|_2^2 \right).
\end{eqnarray*}
Therefore, it implies that
\begin{eqnarray*}
\frac{1-\gamma_2}{2np}  \| \wh\bfa - \bfa_0 \|_2^2  \leq
\frac{3\gamma_1}{2np} \sum_{l\in\calE} \| \bfC_l \bfa_0 \|_2 + \frac{\gamma_2 \| \bfu \|_2^2 }{2np} +\sigma^2 \left[
\frac{1}{n} + \sqrt{\frac{\log (np)}{n^2 p}} \right]  + \frac{1}{n p}
\end{eqnarray*}
holds with probability at least $1-c_3$. \hfill $\blacksquare$

\noindent\textbf{Proof of Theorem \ref{thm:selection_consistency}:}

We only need to prove $P(\|\wh\bfa_p \|_2 =0) \to 1$ by contradiction, the similar arguments apply to $P(\|\wh\bfa_j \|_2 =0) \to 1, j=p_0+1, \ldots, p-1$.

\textbf{\underline{Case 1: $q=1$}}

If $\wh\bfa_p \neq \bfzero$, then $\|\wh\bfa_p \|_2$ is differentiable with respect to its components. The KKT condition for $\bfa_p$ implies
\begin{eqnarray*}
-(\bfx_p - \wh\bfa_p) + \gamma_1 \sum_{l\in\calE} {\rm sgn} \big( (\bfe_{i_1}-\bfe_{i_2})\trans \wh\bfa_p\big)(\bfe_{i_1}-\bfe_{i_2}) + \gamma_2 u_p \frac{\wh\bfa_p}{\|\wh\bfa_p\|_2} =\bfzero
\end{eqnarray*}
Then it follows
\begin{eqnarray*}
\frac{1}{\sqrt{n}} (\wh\bfa_p -\bfa_{0p}) - \frac{1}{\sqrt{n}}\bvarepsilon_p+  \frac{\gamma_1}{\sqrt{n}} \sum_{l\in\calE} {\rm sgn} \big( (\bfe_{i_1}-\bfe_{i_2})\trans \wh\bfa_p\big)(\bfe_{i_1}-\bfe_{i_2}) + \frac{\gamma_2 u_p}{\sqrt{n}} \frac{\wh\bfa_p}{\|\wh\bfa_p\|_2} =\bfzero.
\end{eqnarray*}

Under the conditions for $\gamma_1$ and $\gamma_2$, Theorem \ref{thm:bound1} and remarks thereafter, $\bfa$ is prediction consistent, and thus $\frac{1}{np} \sum_{j=1}^p \| \wh\bfa_j - \bfa_{0j} \|_2^2 =o_P(1)$. It implies with probability to 1, each $\frac{1}{n}\| \wh\bfa_j - \bfa_{0j} \|_2^2 =o_P(1)$. Then the first term is of the order $o_P(1)$. The second term is of the order $o_P(1)$ because $\bvarepsilon_p$ follows a sub-Gaussian distribution. $\|\gamma_2 u_p \wh\bfa_p \|_2 / (\sqrt{n}\|\wh\bfa_p\|_2)= \gamma_2 u_p /\sqrt{n} \to 0$, and hence the forth term is of the order $o_P(1)$. Without loss of the generality, we assume the first entry of $\wh\bfa_p$ is none-zero, i.e., $\wh{A}_{1p}\neq 0$. Also we know $\#\{i | \wh{A}_{ip}\neq \wh{A}_{1p}, i=1,\ldots,n\}/n$ is bounded away from 0. Thus, the first entry of $\sum_{l\in\calE} {\rm sgn} \big( (\bfe_{i_1}-\bfe_{i_2})\trans \wh\bfa_p\big)(\bfe_{i_1}-\bfe_{i_2})$ is of the order $O_P(n)$. Based on the condition $\gamma_1 > 4 \sigma \sqrt{\frac{\log (p \cdot \binom{n}{2})}{n}}$, we know the third term diverges to infinity and dominates other three items, which leads to a contradiction of the KKT condition. Therefore, $\wh\bfa_p=\bfzero$ with a probability tending to one. \hfill $\blacksquare$

\textbf{\underline{Case 2: $q=2$}}

If $\wh\bfa_p \neq \bfzero$, then $\|\wh\bfa_p \|_2$ is differentiable with respect to its components. The KKT condition for $\bfa_p$ implies
\begin{eqnarray*}
-(\bfx_p - \wh\bfa_p) + \gamma_1 \sum_{l\in\calE} \frac{\bfe_{i_1}-\bfe_{i_2}}{\| (\bfI_p \otimes (\bfe_{i_1}-\bfe_{i_2})) \wh\bfa\|_2} + \gamma_2 u_p \frac{\wh\bfa_p}{\|\wh\bfa_p\|_2} =\bfzero
\end{eqnarray*}
Then it follows
\begin{eqnarray*}
\frac{1}{\sqrt{n}} (\wh\bfa_p -\bfa_{0p}) - \frac{1}{\sqrt{n}}\bvarepsilon_p+  \frac{\gamma_1}{\sqrt{n}}\sum_{l\in\calE} \frac{\bfe_{i_1}-\bfe_{i_2}}{\| (\bfI_p \otimes (\bfe_{i_1}-\bfe_{i_2})) \wh\bfa\|_2} + \frac{\gamma_2 u_p}{\sqrt{n}} \frac{\wh\bfa_p}{\|\wh\bfa_p\|_2} =\bfzero.
\end{eqnarray*}

Under the conditions for $\gamma_1$ and $\gamma_2$, Theorem \ref{thm:bound2} and remarks thereafter, $\bfa$ is prediction consistent. Similar with the arguments with the case when $q=1$ above, the first term is of the order $o_P(1)$. The second term is of the order $o_P(1)$ because $\bvarepsilon_p$ follows a sub-Gaussian distribution. $\|\gamma_2 u_p \wh\bfa_p \|_2 / (\sqrt{n}\|\wh\bfa_p\|_2)= \gamma_2 u_p / \sqrt{n} \to 0$, and hence the forth term is of the order $o_P(1)$.

Note that all entries of $\wh\bfa$ cannot the same, and hence $0< \| (\bfI_p \otimes (\bfe_{i_1}-\bfe_{i_2})) \wh\bfa\|_2 \leq \| (\bfI_p \otimes (\bfe_{i_1}-\bfe_{i_2}))\|_2 \|\wh\bfa\|_2 \leq \| (\bfI_p \otimes (\bfe_{i_1}-\bfe_{i_2}))\|_F \|\wh\bfa\|_2 \leq \sqrt{2p}\|\wh\bfa\|_2$. Also, we have $1/\|\wh\bfa \|_2=O_P(\sqrt{np})$. Thus, the first entry of $\sum_{l\in\calE} \frac{\bfe_{i_1}-\bfe_{i_2}}{\| (\bfI_p \otimes (\bfe_{i_1}-\bfe_{i_2})) \wh\bfa\|_2}$ is of the order $O_P(n^{\frac{3}{2}})$. Based on the condition $\gamma_1 > 4 \sigma \sqrt{\frac{\log (p \cdot \binom{n}{2})}{n}}$, we know the third term diverges to infinity and dominates other three items, which leads to a contradiction of the KKT condition. Therefore, $\wh\bfa_p=\bfzero$ with a probability tending to one. \hfill $\blacksquare$

\subsection*{S.3 Degrees of freedom}
\label{sec:degree}

In this section, we provide unbiased estimators for the degrees of freedoms of sparse convex clustering. Degrees of freedom is generally defined in regression problems to explain the amount of flexibility in the model. It is a key component for model selection and statistical hypothesis testing. Note that our sparse convex clustering can be formulated as a penalized
regression problem for which the degrees of freedom can be established.
Motivated by \citet{Tan2015}, we develop unbiased estimators for
the degrees of freedom of sparse convex clustering with $q=1$ in
Lemma \ref{thm:df1} and $q=2$ in Lemma \ref{thm:df2}.
For simplicity, we consider the case with $w_l=1, l \in \calE$ in the following theoretical developments.

\begin{lemma} \label{thm:df1} Assume $X_{i\cdot}\stackrel{iid}{\sim}\textrm{MVN}_{p}(\bmu,\sigma^{2}\bfI_{p})$, and let $\widehat{\bfa} \buildrel \Delta \over = \textrm{vec}(\wh{\bfA})$
be the solution to $(\ref{eq:obj_constraint})$ with $q=1$. Then we
have ${\rm df}\buildrel \Delta \over ={\rm tr}(\frac{\partial\wh\bfa}{\partial\bfx})$
is of the form
\begin{eqnarray*}
{\rm df}_{1} & = & {\rm tr}\Bigg(\Bigg[\bfI+\gamma_{2}\bfP_{1}\sum_{s\in\calB_{12}}\bigg(\frac{\bfD_{s}\trans\bfD_{s}} {\|\bfD_{s}\wh\bfa\|_{2}}-\frac{\bfD_{s}\trans\bfD_{s}\wh\bfa\wh\bfa\trans\bfD_{s}\trans\bfD_{s}}{\|\bfD_{s}\wh\bfa\|_{2}^{3}}\bigg)\Bigg]^{-1}\bfP_{1}\Bigg),
\end{eqnarray*}
where $\bfD_{s}$ and $\bfP_{1}$ are defined in $(\ref{eqn:defD})$ and $(\ref{eqn:defP})$, respectively.
\end{lemma}

Following a similar proof technique, we provide an unbiased estimator for the degrees of freedom of the sparse convex clustering with $q=2$.

\begin{lemma} \label{thm:df2} Assume $X_{i\cdot}\stackrel{iid}{\sim}\textrm{MVN}_{p}(\bmu,\sigma^{2}\bfI_{p})$, and let $\widehat{\bfa}$
be the solution to $(\ref{eq:obj_constraint})$ with $q=2$. Therefore,
the degrees of freedom is
\begin{eqnarray*}
{\rm df}_{2} & = & {\rm tr}\Bigg(\Bigg[\bfI+\gamma_{1}\bfP_{2}\sum_{s\in\wh{\calB}_{2}\cap\{1,\ldots,|\calE|\}}\left(\frac{\bfD_{s}\trans\bfD_{s}}{\|\bfD_{s}\wh\bfa\|_{2}}-\frac{\bfD_{s}\trans\bfD_{s}\wh\bfa\wh{\bfa}\trans\bfD_{s}\trans\bfD_{s}}{\|\bfD_{s}\wh\bfa\|_{2}^{3}}\right)\\
 &  & +\gamma_{2}\bfP_{2}\sum_{s\in\wh{\calB}_{2}\cap\{|\calE|+1,\ldots,|\calE|+p\}}\left(\frac{\bfD_{s}\trans\bfD_{s}}{\|\bfD_{s}\wh\bfa\|_{2}}-\frac{\bfD_{s}\trans\bfD_{s}\wh\bfa\wh{\bfa}\trans\bfD_{s}\trans\bfD_{s}}{\|\bfD_{s}\wh\bfa\|_{2}^{3}}\right)\Bigg]^{-1}\bfP_{2}\Bigg).
\end{eqnarray*}
\end{lemma}
\textbf{Proofs of Lemmas \ref{thm:df1}-\ref{thm:df2}:} Following the arguments in \citet{Tan2015}, the number of degrees of freedom
(df) of \eqref{eq:obj_constraint}, when $q=1$ or $2$, can be derived
under the assumption $w_{l}=1,l\in\calE$ and $X_{i\cdot}\sim\textrm{MVN}_{p}(\bmu,\sigma^{2}\bfI_{p})$.

\underline{\textbf{Case $q=1$:}} Rewrite \eqref{eq:obj_constraint}
into the following formulation:
\begin{eqnarray}
\min_{\bfa\in\mathbb{R}^{np\times1}} &  & \frac{1}{2}\|\bfx-\bfa\|_{2}^{2}+\gamma_{1}\sum_{l\in\calE}w_{l}\|\bfC_{l}\bfa\|_{1}+\gamma_{2}\sum_{j=1}^{p}u_{j}\|(\bfe_{j}^{*T}\otimes\bfI_{n})\bfa\|_{2},\label{obj_df_1}
\end{eqnarray}
where $\bfe_{j}^{*}$ is a $p$-dim vector with its $j$-th element
as 1 and 0 otherwise.

Define
\begin{equation}
\bfD_{j} = \begin{cases}  w_{j}\bfC_{j} & \mbox{if } s=1,\ldots,|\calE| \\ u_{s-|\calE|}(\bfe_{s-|\calE|}^{*t}\otimes\bfI_{n}), & \mbox{if } s=|\calE|+1,\ldots,|\calE|+p \end{cases}, \label{eqn:defD}
\end{equation}
and let $\bfD\trans=(\bfD_{1}\trans,\ldots,\bfD_{|\calE|+p}\trans)$. Then \eqref{obj_df_1} becomes
\begin{eqnarray*}
\min_{\bfa\in\mathbb{R}^{np\times1}} &  & \frac{1}{2}\|\bfx-\bfa\|_{2}^{2}+\gamma_{1}\sum_{s=1}^{|\calE|}\|\bfD_{s}\bfa\|_{1}+\gamma_{2}\sum_{s=|\calE|+1}^{|\calE|+p}\|\bfD_{s}\bfa\|_{2}.
\end{eqnarray*}
Actually, the second term can be written component-wisely into $\gamma_{1}\sum_{s=1}^{|\calE|}\sum_{j=1}^{p}|\bfd_{sj}\trans\bfa|$,
where $\bfd_{sj}$ is the vector consisting of the $j$-th row of
$\bfD_{s}$.

Let $\wh{\calB}_{1}=\wh{\calB}_{11}\bigcup\wh{\calB}_{12}$, where
$\wh{\calB}_{11}=\{(s,j):|\bfd_{sj}\trans\wh\bfa|\neq0,s=1,\ldots,|\calE|,j=1,\ldots,p\}$
and $\wh{\calB}_{12}=\{s:\|\bfD_{s}\wh{\bfa}\|_{2}\neq0,s=|\calE|+1,\ldots,|\calE|+p\}$.
The derivative of \eqref{obj_df_1} is obtained as
\begin{eqnarray*}
\bfx-\bfa=\gamma_{1}\sum_{s=1}^{|\calE|}\sum_{j=1}^{p}f_{sj}\bfd_{sj}+\gamma_{2}\sum_{s=|\calE|+1}^{|\calE|+p}\bfD_{s}\trans\bfg_{s},
\end{eqnarray*}
where $f_{sj}=\textrm{sgn}(\bfd_{sj}\trans\wh{\bfa})$, if $(s,j)\in\wh{\calB}_{11}$
and $f_{sj}\in[-1,1]$, if $s\notin\wh{\calB}_{11}$, and $\bfg_{s}={\bfD_{s}\wh\bfa}/{\|\bfD_{s}\wh\bfa\|_{2}}$,
if $s\in\wh{\calB}_{12}$ and $\bfg_{s}\in\{\Gamma:\|\Gamma\|_{2}\leq1\}$,
if $s\notin\wh{\calB}_{12}$.

Define matrix $\bfD_{-\wh{\calB}_{1}}$ by removing the rows
of $\bfD$ corresponding to those elements in $\wh{\calB}_{1}$, and
\begin{equation}
\bfP_{1}=\bfI-\bfD_{-\wh{\calB}_{1}}\trans(\bfD_{-\wh{\calB}_{1}}\bfD_{-\wh{\calB}_{1}}\trans)^{+}\bfD_{-\wh{\calB}_{1}}. \label{eqn:defP}
\end{equation}
By the property $\bfD_{-\wh{\calB}_{1}}\wh\bfa=\bfzero$,
\begin{eqnarray*}
\bfP_{1}\bfx-\wh\bfa & = & \gamma_{1}\bfP_{1}\sum_{s=1}^{|\calE|}\sum_{j=1}^{p}f_{sj}\bfd_{sj}+\gamma_{2}\bfP_{1}\sum_{s=|\calE|+1}^{|\calE|+p}\bfD_{s}\trans\bfg_{s}\\
 & = & \gamma_{1}\bfP_{1}\sum_{(s,j)\in\calB_{11}}\textrm{sgn}(\bfd_{sj}\trans\wh{\bfa})\bfd_{sj}+\gamma_{2}\bfP_{1}\sum_{s\in\calB_{12}}\frac{\bfD_{s}\trans\bfD_{s}\wh\bfa}{\|\bfD_{s}\wh\bfa\|_{2}}.
\end{eqnarray*}

By the property shown by \citet{Vaiter2012}; i.e., there exists a
neighborhood around almost every $\bfx$ such that the solution $\wh\bfa$
is locally constant with respect to $\bfx$, the derivative of the
above equation with respect to $\bfx$ is
\begin{eqnarray*}
\bfP_{1}-\frac{\partial\wh\bfa}{\partial\bfx} & = & \gamma_{2}\bfP_{1}\sum_{s\in\calB_{12}}\left(\frac{\bfD_{s}\trans\bfD_{s}}{\|\bfD_{s}\wh\bfa\|_{2}}-\frac{\bfD_{s}\trans\bfD_{s}\wh\bfa\wh\bfa\trans\bfD_{s}\trans\bfD_{s}}{\|\bfD_{s}\wh\bfa\|_{2}^{3}}\right)\frac{\partial\wh\bfa}{\partial\bfx}.
\end{eqnarray*}
Therefore, ${\rm df}\triangleq\textrm{tr}(\frac{\partial\wh\bfa}{\partial\bfx})$
is of the form
\begin{eqnarray*}
\textrm{df}_{1} & = & \textrm{tr}\Bigg(\Bigg[\bfI+\gamma_{2}\bfP_{1}\sum_{s\in\calB_{12}}\bigg(\frac{\bfD_{s}\trans\bfD_{s}}{\|\bfD_{s}\wh\bfa\|_{2}}-\frac{\bfD_{s}\trans\bfD_{s}\wh\bfa\wh\bfa\trans\bfD_{s}\trans\bfD_{s}}{\|\bfD_{s}\wh\bfa\|_{2}^{3}}\bigg)\Bigg]^{-1}\bfP_{1}\Bigg).
\end{eqnarray*}

\underline{\textbf{Case $q=2$:}} Rewrite \eqref{eq:obj_constraint}
into the following form when $q=2$, the $L_{2}$-norm:
\begin{eqnarray}
\min_{\bfa\in\mathbb{R}^{np\times1}} &  & \frac{1}{2}\|\bfx-\bfa\|_{2}^{2}+\gamma_{1}\sum_{j=1}^{|\calE|}\|\bfD_{s}\bfa\|_{2}+\gamma_{2}\sum_{s=|\calE|+1}^{|\calE|+p}\|\bfD_{s}\bfa\|_{2}.\label{new_obj_df_2}
\end{eqnarray}

Let $\wh{\calB}_{2}=\{s:\|\bfD_{s}\wh{\bfa}\|_{2}\neq0,s=1,\ldots,|\calE|+p\}$.
The derivative of \eqref{new_obj_df_2} is obtained as
\begin{eqnarray*}
\bfx-\bfa=\gamma_{1}\sum_{s=1}^{|\calE|}\bfD_{s}\trans\bfg_{j}+\gamma_{2}\sum_{s=|\calE|+1}^{|\calE|+p}\bfD_{s}\trans\bfg_{j},
\end{eqnarray*}
where $\bfg_{s}={\bfD_{s}\wh\bfa}/{\|\bfD_{s}\wh\bfa\|_{2}}$, if
$s\in\wh{\calB}_{2}$ and $\bfg_{s}\in\{\Gamma:\|\Gamma\|_{2}\leq1\}$,
if $s\notin\wh{\calB}_{2}$. Then, define matrix $\bfD_{-\wh{\calB}_{2}}$
by removing the rows of $\bfD$ corresponding to those elements in
$\wh{\calB}_{2}$, and $\bfP_{2}=\bfI-\bfD_{-\wh{\calB}_{2}}\trans(\bfD_{-\wh{\calB}_{2}}\bfD_{-\wh{\calB}_{2}}\trans)^{+}\bfD_{-\wh{\calB}_{2}}$.
It follows
\begin{eqnarray*}
\frac{\partial\wh\bfa}{\partial\bfx} & = & \Bigg[\bfI+\gamma_{1}\bfP_{2}\sum_{s\in\wh{\calB}_{2}\cap\{1,\ldots,|\calE|\}}\left(\frac{\bfD_{s}\trans\bfD_{s}}{\|\bfD_{s}\wh\bfa\|_{2}}-\frac{\bfD_{s}\trans\bfD_{s}\wh\bfa\wh{\bfa}\trans\bfD_{s}\trans\bfD_{s}}{\|\bfD_{s}\wh\bfa\|_{2}^{3}}\right)\\
 &  & +\gamma_{2}\bfP_{2}\sum_{s\in\wh{\calB}_{2}\cap\{|\calE|+1,\ldots,|\calE|+p\}}\left(\frac{\bfD_{s}\trans\bfD_{s}}{\|\bfD_{s}\wh\bfa\|_{2}}-\frac{\bfD_{s}\trans\bfD_{s}\wh\bfa\wh{\bfa}\trans\bfD_{s}\trans\bfD_{s}}{\|\bfD_{s}\wh\bfa\|_{2}^{3}}\right)\Bigg]^{-1}\bfP_{2}.
\end{eqnarray*}
Therefore, the df when $q=2$ is of the form
\begin{eqnarray*}
\text{df}_{2} & = & \text{tr}\Bigg(\Bigg[\bfI+\gamma_{1}\bfP_{2}\sum_{s\in\wh{\calB}_{2}\cap\{1,\ldots,|\calE|\}}\left(\frac{\bfD_{s}\trans\bfD_{s}}{\|\bfD_{s}\wh\bfa\|_{2}}-\frac{\bfD_{s}\trans\bfD_{s}\wh\bfa\wh{\bfa}\trans\bfD_{s}\trans\bfD_{s}}{\|\bfD_{s}\wh\bfa\|_{2}^{3}}\right)\\
 &  & +\gamma_{2}\bfP_{2}\sum_{s\in\wh{\calB}_{2}\cap\{|\calE|+1,\ldots,|\calE|+p\}}\left(\frac{\bfD_{s}\trans\bfD_{s}}{\|\bfD_{s}\wh\bfa\|_{2}}-\frac{\bfD_{s}\trans\bfD_{s}\wh\bfa\wh{\bfa}\trans\bfD_{s}\trans\bfD_{s}}{\|\bfD_{s}\wh\bfa\|_{2}^{3}}\right)\Bigg]^{-1}\bfP_{2}\Bigg).
\end{eqnarray*}


\end{document}